\documentclass{aa}
\pdfoutput=1
\usepackage{txfonts} 

\usepackage{graphicx}
\usepackage{natbib} 
\bibpunct{(}{)}{;}{a}{}{,} 
\usepackage{color}

\title{Large dust gaps in the transitional disks of HD\,100453 and HD\,34282}
\subtitle{Connecting the gap size to the spectral energy distribution and mid-infrared imaging}
\author{S. Khalafinejad\inst{1,2} \and K. M. Maaskant\inst{2} \and N. Mari\~nas\inst{3} \and A.G.G.M Tielens\inst{2} } 

\institute{Hamburg Observatory, Hamburg University, Gojenbergsweg 112, 21029 Hamburg, Germany \and Leiden Observatory, Leiden University, P.O. Box 9513, 2300 RA Leiden, The Netherlands \and Astronomy Department, University of Florida, Gainesville, FL 32611, USA }

\abstract{The formation of dust gaps in protoplanetary disks is one of the most important signposts of disk evolution and possibly the formation of planets.}
{We aim to characterize the `flaring' disk structure around the Herbig Ae/Be stars HD~100453 and HD~34282. Their spectral energy distributions (SEDs) show an emission excess between $15-40$ $\mu$m, but very weak (HD~100453) and no (HD~34282) signs of the 10 and 20 $\mu$m amorphous silicate features. We investigate whether this implies the presence of large dust gaps. }
{Spatially resolved mid-infrared Q-band images taken with Gemini North/MICHELLE are investigated. We perform radiative transfer modeling and examine the radial distribution of dust. We simultaneously fit the Q-band images and SEDs of HD~100453 and HD~34282.}
{Our solutions require that the inner-halos and outer-disks are likely separated by large dust gaps that are depleted with respect to the outer disk by a factor of 1000 or more. The inner edges of the outer disks of HD~100453 and HD~34282 have temperatures of $\sim160 \pm 10$ K and $\sim60 \pm 5$ K respectively. Because of the high surface brightnesses of these walls, they dominate the emission in the Q-band. Their radii are constrained at $20^{+2}_{-2}$ AU and $92^{+31} _{-17}$ AU, respectively.}
{HD~100453 and HD~34282 likely have disk dust gaps and the upper limit on the dust mass in each gap is estimated to be about 10$^{-7}$M$_{\odot}$.  We find that the locations and sizes of disk dust gaps are connected to the SED, as traced by the mid-infrared flux ratio $F_{30}/F_{13.5}$. We propose a new classification scheme for the Meeus groups \citep{2001Meeus} based on the $F_{30}/F_{13.5}$ ratio. The absence of amorphous silicate features in the observed SEDs is caused by the depletion of small ($\lesssim1$~$\mu$m) silicate dust at temperatures above $\gtrsim160$ K, which could be related to the presence of a dust gap in that region of the disk.}


\date{\today} 

\begin{document}

\maketitle

\section{Introduction}

Transitional disks have dust depleted gaps and inner holes in their dust distribution and form a special class of protoplanetary disks \citep{2011WilliamsCieza}. The presence of dust gaps and inner dust holes may be indicators that planets are forming in the disks. To search for evidence of planet formation and characterize their physical and chemical conditions, the location and sizes of dust gaps in protoplanetary disks have to be investigated. 

Transitional disks can be identified on the basis of their low near infrared excess (e.g. \citealt{2005Calvet, 2007Espaillat, 2007Najita}). Analysis of their spectral energy distribution (SED) may indicate that their inner regions are depleted of dust by several orders of magnitude (See Figure \ref{fig:surface_density}, section 4 and \citealp{2013Maaskant}). However, modeling the radial disk structure by fitting only the SED is highly degenerate. Dust imaging of a protoplanetary disks enables to study the spatial distribution of dust grains of different sizes in the disk. High spatial resolution observations of transitional disks  reveal complex disk structures and can be used to study the interaction between dust gaps and proto-planets (e.g.  \citealt{2013vanderMarel,2013Casassus,2013Quanz}). Characterizing the connection between the radial structure and the SED is thus important to gain insight in the role of planet formation in the evolution of protoplanetary disks.

\begin{figure*}[t]
\centering
\includegraphics[width=0.75\textwidth]{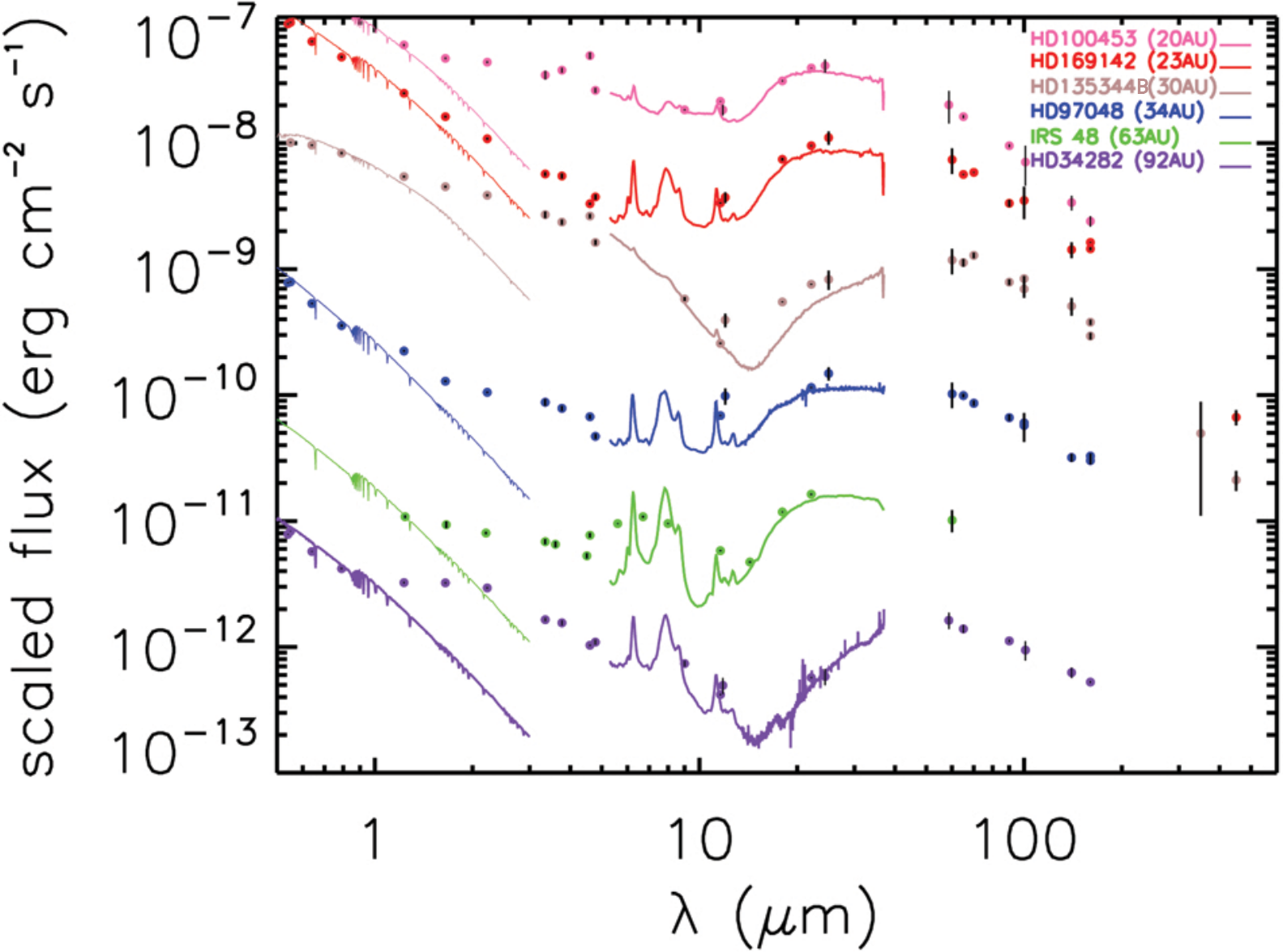}
\includegraphics[width=0.18\textwidth]{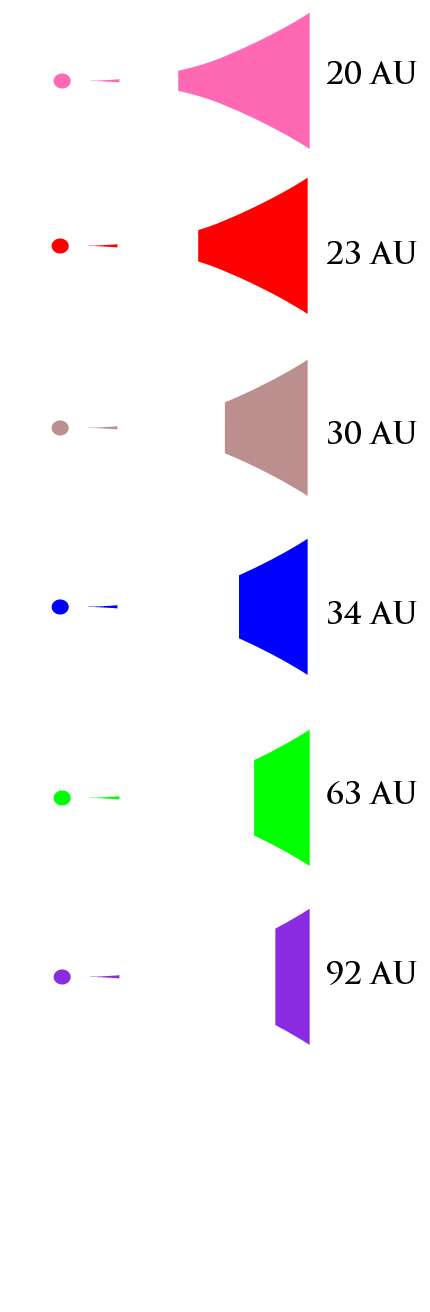}

\caption{\label{fig:all} The SEDs of group Ib Herbig Ae/Be stars in the sample of \citet{2014Maaskant}. There is a great similarity in the shape of the SEDs. The fluxes are scaled so that the objects are sorted by dust gap size, with the smallest dust gap radius on top and the largest dust gap radius on the bottom. The flux scaling factors for the objects from top to bottom are respectively 10, 5, 1, 0.03, 0.003, 0.003.  }
\end{figure*}

Herbig Ae/Be stars are intermediate mass stars with circumstellar disks (e.g. \citealt{2007Natta}). The SEDs of Herbig Ae/Be stars fall apart into two groups \citep{2001Meeus}. Group I, with strong excess at mid- to far-infrared wavelengths, and group II, without strong excess at mid- to far-infrared, but with a remarkable similarity in spectral shape. A first interpretation for the evolutionary link between these groups was proposed by \citet{2004bDullemond, 2005Dullemond}. These authors suggested that grain growth and settling cause the mid- to far-infrared excess to decrease. It was proposed that in this scenario, the disk structure evolves from flaring (group I) to flat (group II). However, recent studies indicate that almost all group I objects have large dust gaps \citep{2005Grady, 2012Honda, 2013Maaskant}. This implies that it is unlikely that group I sources with large dust gaps can evolve to group II disks, where no large dust gaps are found. To solve this issue, \citet{2013Maaskant} suggested that both groups evolve from a primordial continuous flaring disk, but may follow different pathways. The disks of group I objects are flaring/transitional due to the formation of large dust gaps. The disks of group II objects are self-shadowed because grain growth and vertical settling have flattened the outer disk. 
In addition to the geometrical classification, the flaring and flat disks with silicate features are respectively called group Ia and group IIa; those without silicate features are called group Ib and group IIb.

In this paper, we investigate spatially resolved direct imaging mid-infrared observations of two protoplanetary disks presented in \citet{2011Marinas}. As these images are most sensitive to thermal emission of micron sized grains, they are suitable to study the radial density structure of transitional disks with large dust depleted gaps. To derive the properties of the disk, we perform a similar analysis of Q-band images as carried out in \citet{2013Maaskant}.

The content of this paper is outlined in the following way. In Section~\ref{sec:thesample}, we introduce the Herbig stars HD~100453 and HD~34282. In Section~\ref{sec:observations} we discuss the Q-band observations and photometric properties of our sample. Section \ref{sec:modeling} describes the radiative transfer code MCMax and the dust model. In section~\ref{sec:results} we derive the properties of the disk structures and constrain the radii of the inner edges of the outer disks.  The discussion and conclusions are given in Sections~\ref{sec:discussion} and~\ref{sec:conclusions}.

\section{The sample}
\label{sec:thesample}
\begin{figure}[t]
\centering
\includegraphics[width=\columnwidth]{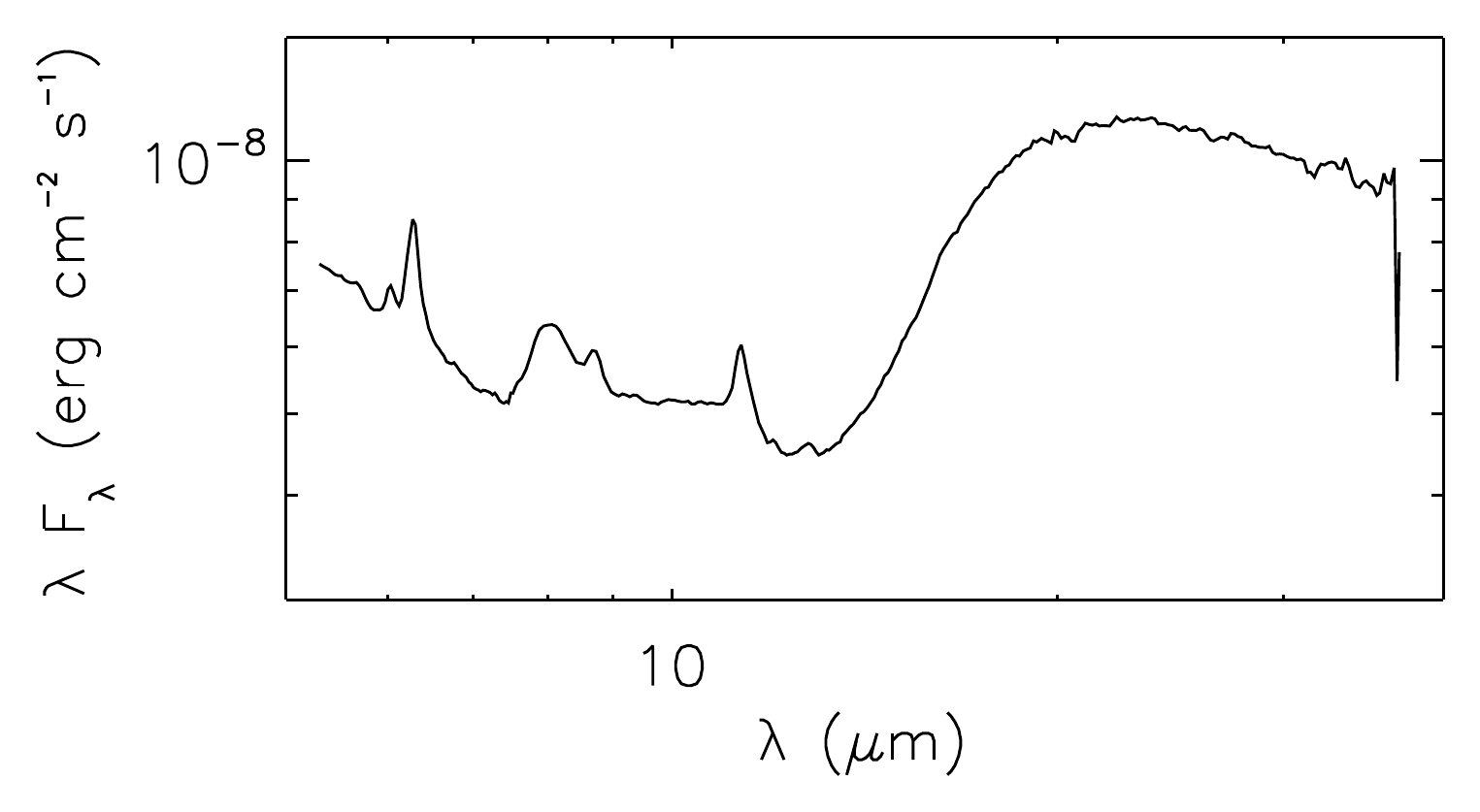}
\caption{\label{fig:100453zoom} A close up of the Spitzer/IRS spectrum of HD~100453 shows a tentative detection of the 10 and 20 $\mu$m silicate features.  }
\end{figure}

The Herbig Ae/Be objects HD~100453 and HD~34282 are studied in this paper. In this section, we present a brief summary. The SEDs of HD~100453 and HD~34282 are characterized by a strong excess of MIR emission at $\sim15-40$ $\mu$m (Figure \ref{fig:all}). HD~100453 may show a very weak sign of amorphous silicate features (Figure \ref{fig:100453zoom}), while in HD~34282 they are totally absent. The spectra of HD~100453 and HD~34282 show PAH emission bands. Their $I_{6.2}/I_{11.3}$ feature ratios are respectively 2.25 and 1.82 and relatively high, possibly indicating optically thin gas flows through the disk dust gaps \citep{2014Maaskant}. 

\subsection{HD~100453}

HD\,100453 is thought to be in transition between a gas rich protoplanetary disk and a gas poor debris disk \citep{2009Collins}.
Observations of spatially resolved Q-band imaging \citep{2011Marinas}, the SED and the absence of the silicate feature are indications of a dust gap \citet{2013Maaskant}. HD 100453 has a close M-type companion (projected distance of 120 AU), and the connection of the companion with the disk structure is not well understood \citep{2009Collins}. In \citealt{2009Collins}, the authors presented two non-detections. The first one is CO 4.97 $\mu$m ro-vibrational emission with Gemini/Pheonix (these transitions trace hot and warm gas), and the second one is CO 3-2 rotational emission with JCMT/HARP at 867 micron (this is a cold-gas tracer). In addition, \citealt{2013Meeus} present the non detections of CO rotational emission in the spectral range of 50 to 210 micron (cold and warm gas). Following the discussion of the non detection of gas tracers in HD~100453, \citealt{2008Carmona} report the non-detection of H$_{2}$ S(1) and S(2) rotational lines at 17 and 12 micron (this traces warm gas at a few hundred Kelvin). However some gas must be present in the disk as the [OI] 63 micron line is detected \citep{2012Meeus,2013Fedele}. The stellar properties are taken from \citet{2001Meeus}, where the distance $114^{+11}_{-9}$ pc is adopted based on Hipparcos measurements. 

\subsection{HD~34282}

\citet{2009Acke} already suggested that the disk around HD\,34282 has a large opacity dust gap based on the near IR excess and far IR color. Rotational J=3-2 CO emission \citep{2000Greaves}, but no rovibrational CO emission \citep{2005Carmona} was detected in this source, which is consistent with an evacuated inner disk but gas rich outer disk. \citet{2005Dent} also report the detection of $^{12}$CO 3-2 emission in HD~34282. They constrain the inclination of HD~34282 to $50\pm5$ degrees.
We adopt the stellar properties derived by \citet{2004Merin} putting this source at a distance of $348^{+129}_{-77}$ pc.

\section{Observations}
\label{sec:observations}
\begin{figure*}
	\centering
\includegraphics[width=0.45\textwidth]{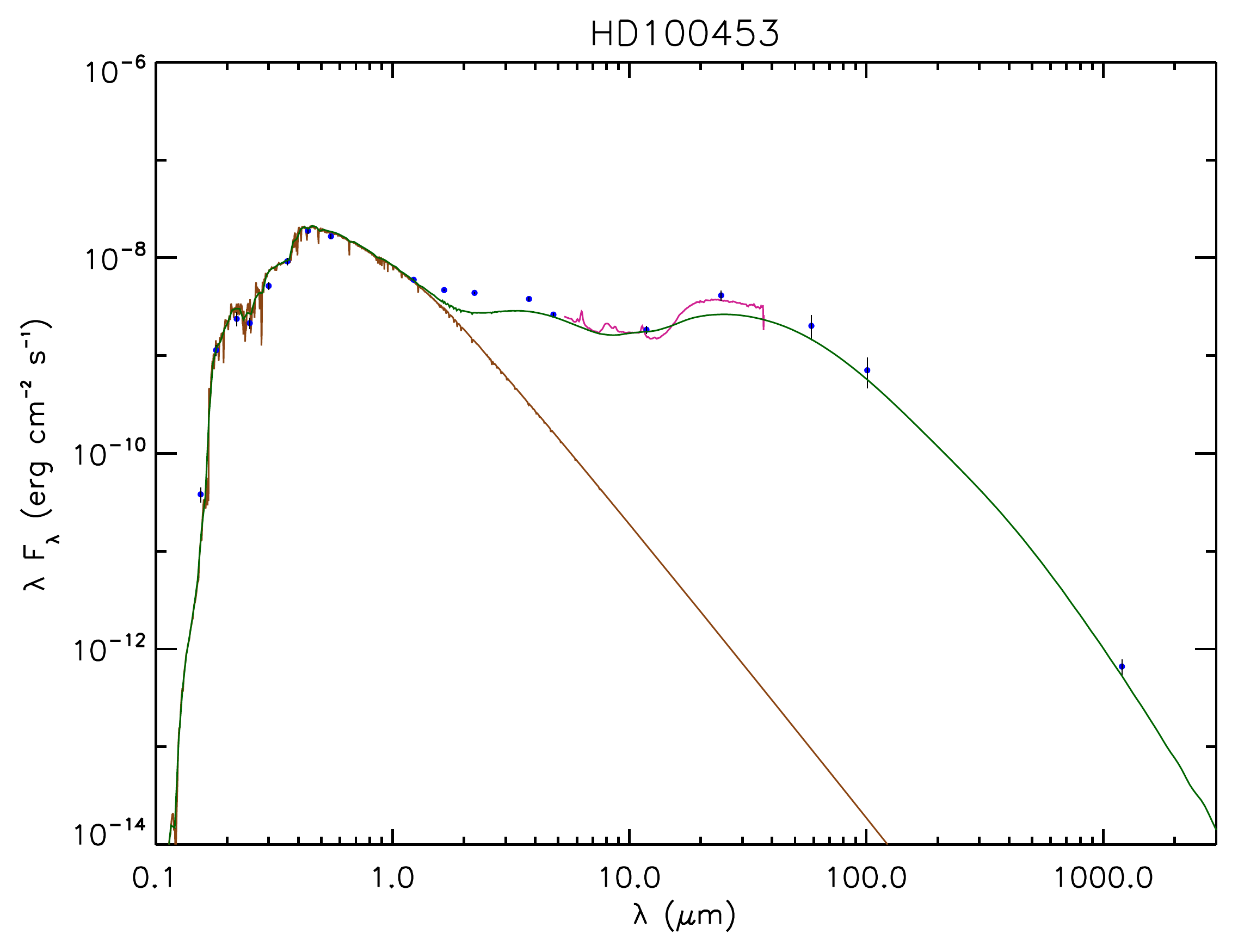}
\includegraphics[width=0.45\textwidth]{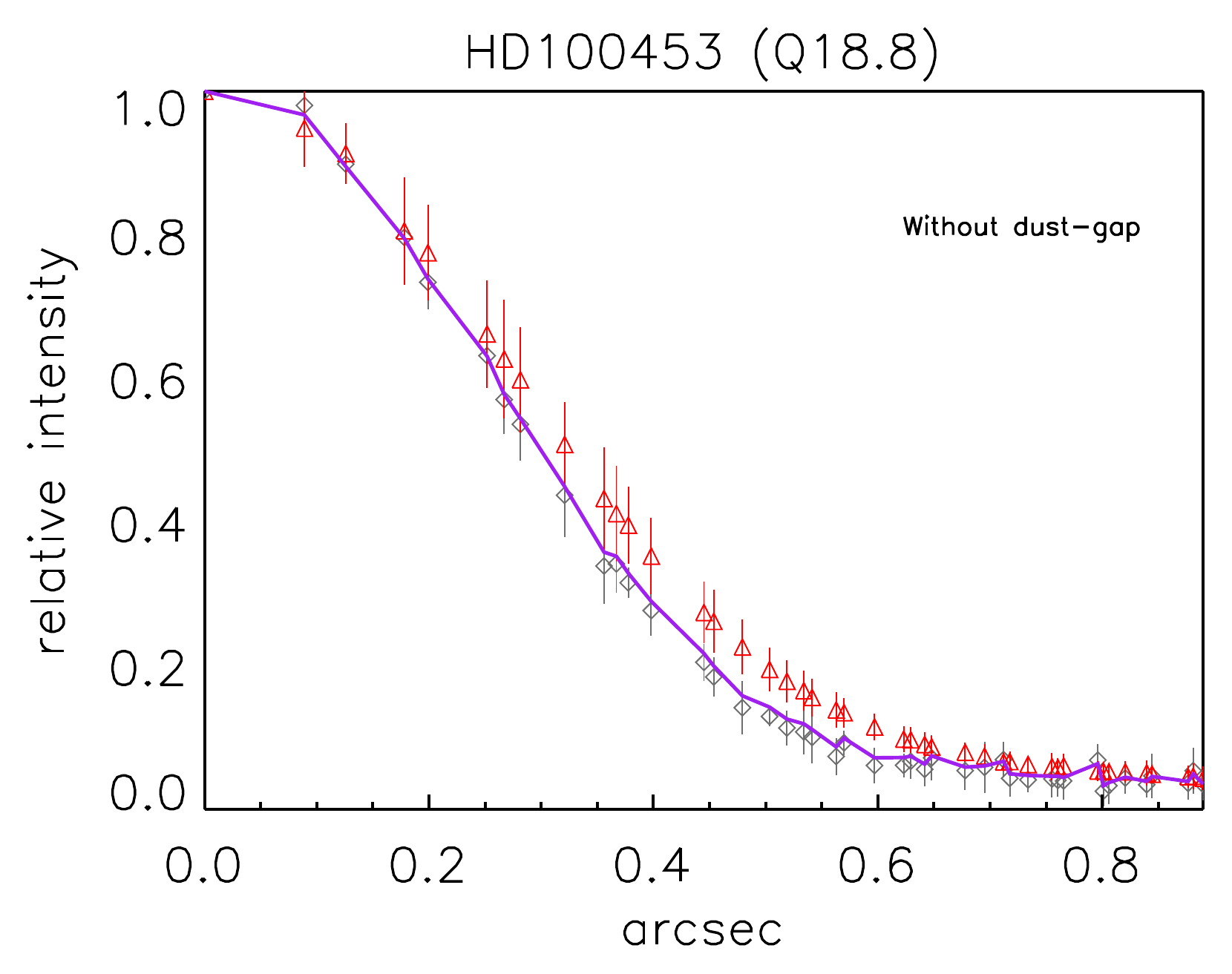}
\includegraphics[width=0.45\textwidth]{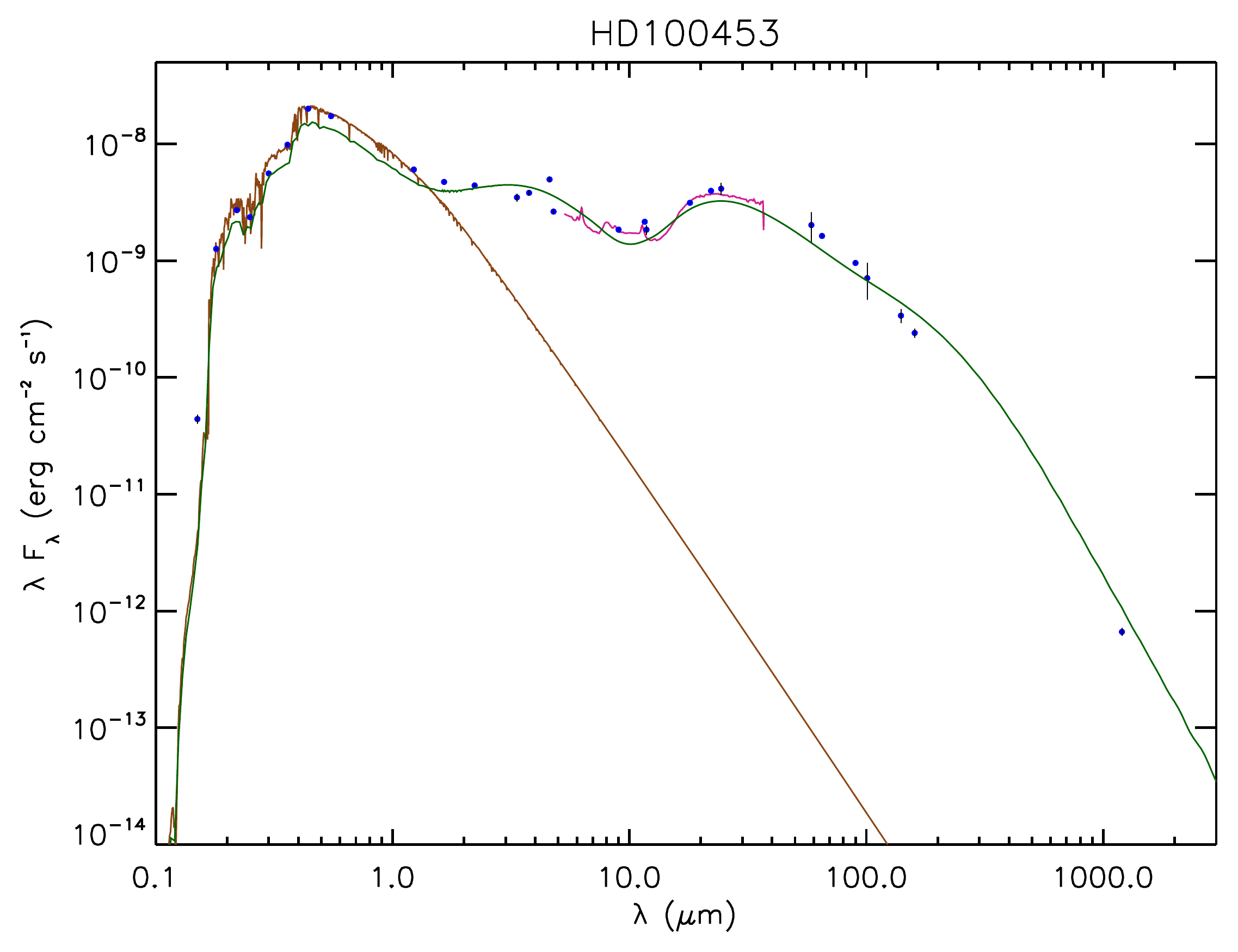}
\includegraphics[width=0.45\textwidth]{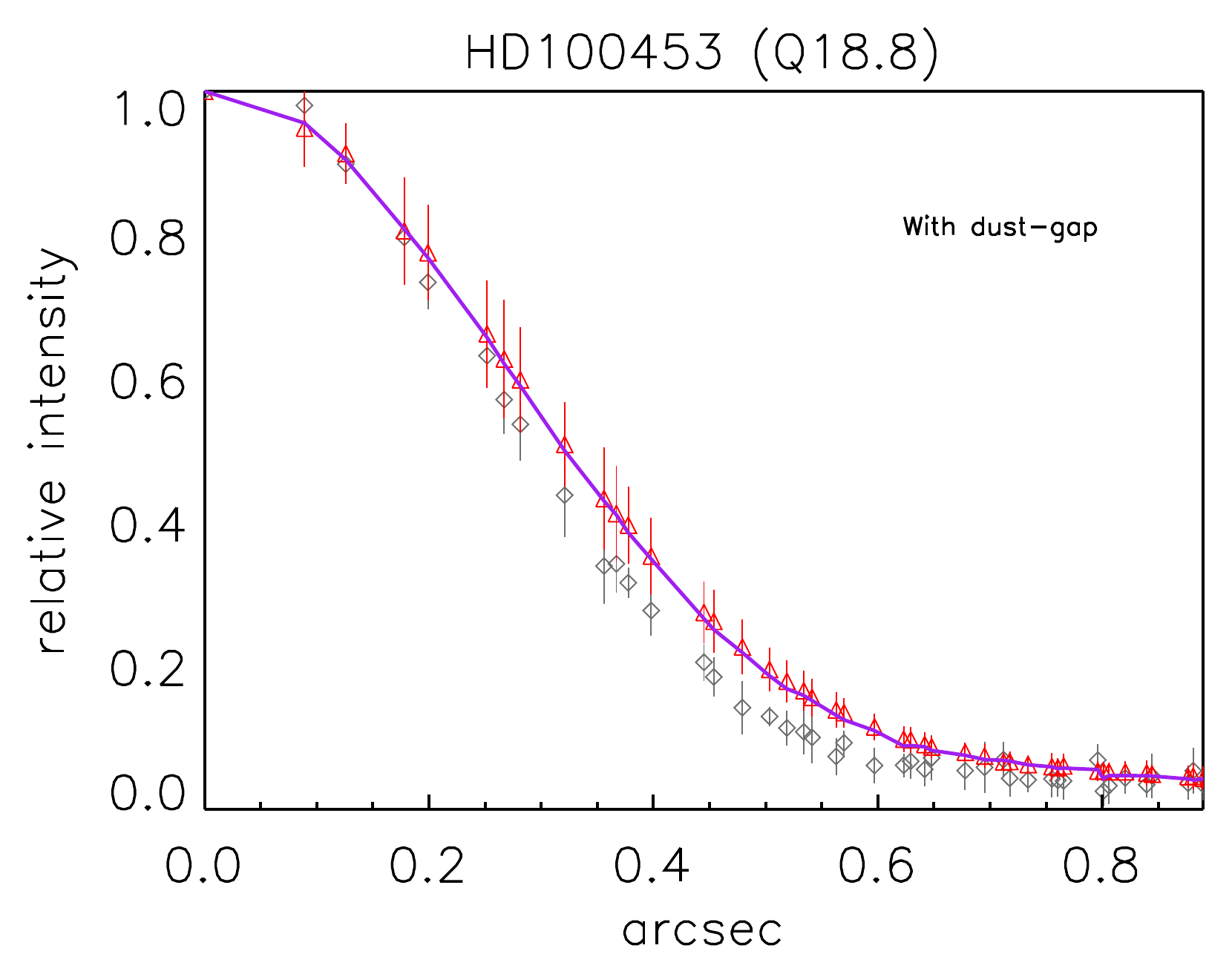}

\caption{ \label{models_hd100453} The continuous disk (top) and transitional disk (bottom) models for HD~100453. \textbf{Left:} the solid brown lines show the stellar Kurucz models. The solid red lines show the Spitzer/IRS spectra. The blue dots represent the observed photometry. The solid green lines show the total fluxes. \textbf{Right:} azimuthally averaged radial brightness profiles of the Q-band relative to the maximum flux. The central wavelengths of these images are $18.8 \mu$m. The grey diamonds indicate the PSF of the calibration star. The red triangles show the observation of the science targets.  The error bars indicate the one sigma variations on the azimuthally averaged radial brightness profiles and are thus a reflection of the source asymmetry. The solid purple line shows our best-fit model. \textbf{Top left:} The model fit to the SED seems reasonable, although the detailed spectral shape at $\sim$20 $\mu$m is not well fitted. \textbf{Top right:} the convolved model image does not fit the observed radial brightness profile.  \textbf{Bottom left:} the assumption of a large gap in the disk gives a better fit to the SED at MIR wavelengths. \textbf{Bottom right:} The disk-gap model fits the extended emission as observed in the Q-band image. }
\end{figure*}

\begin{figure*}
	\centering

\includegraphics[width=0.45\textwidth]{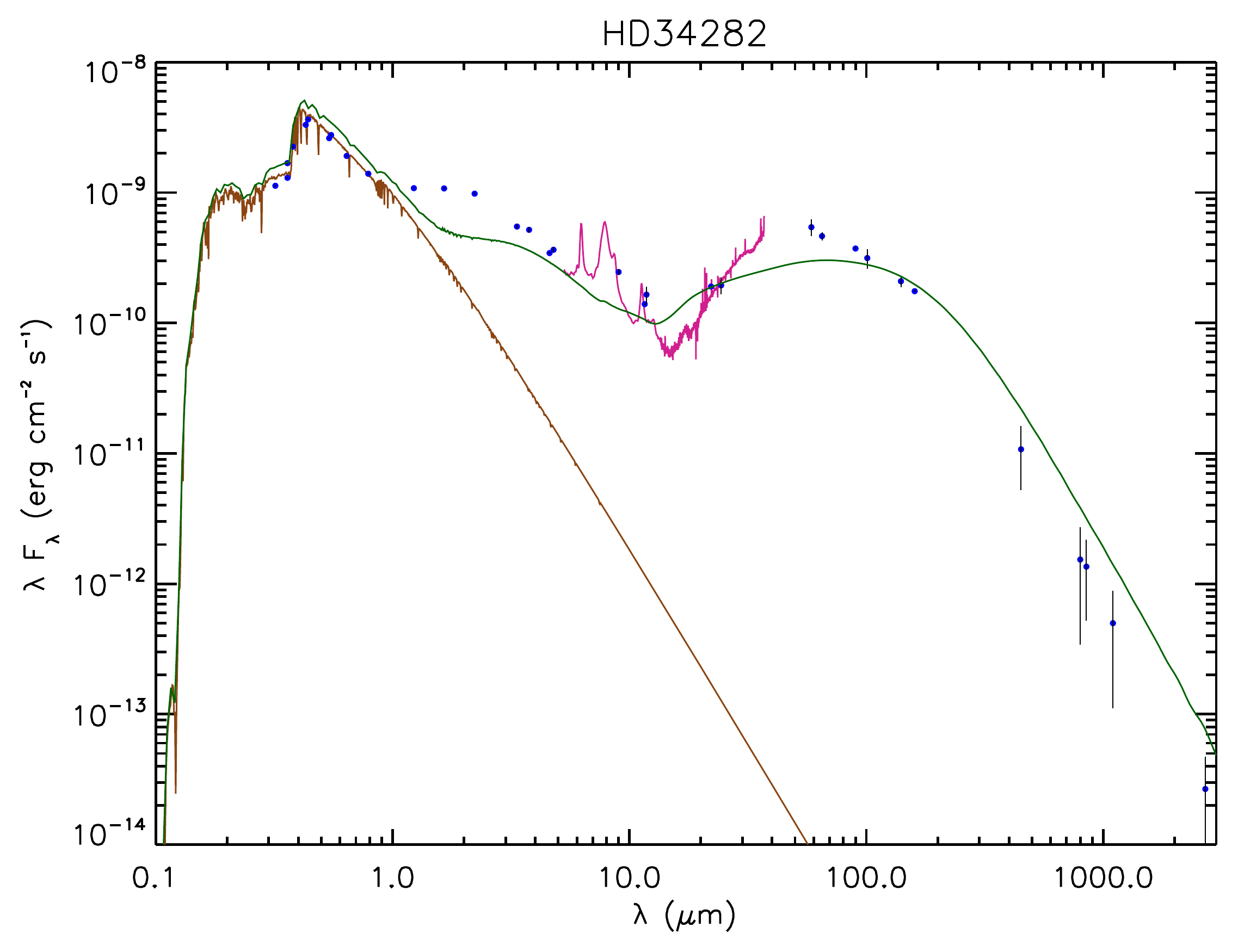}
\includegraphics[width=0.45\textwidth]{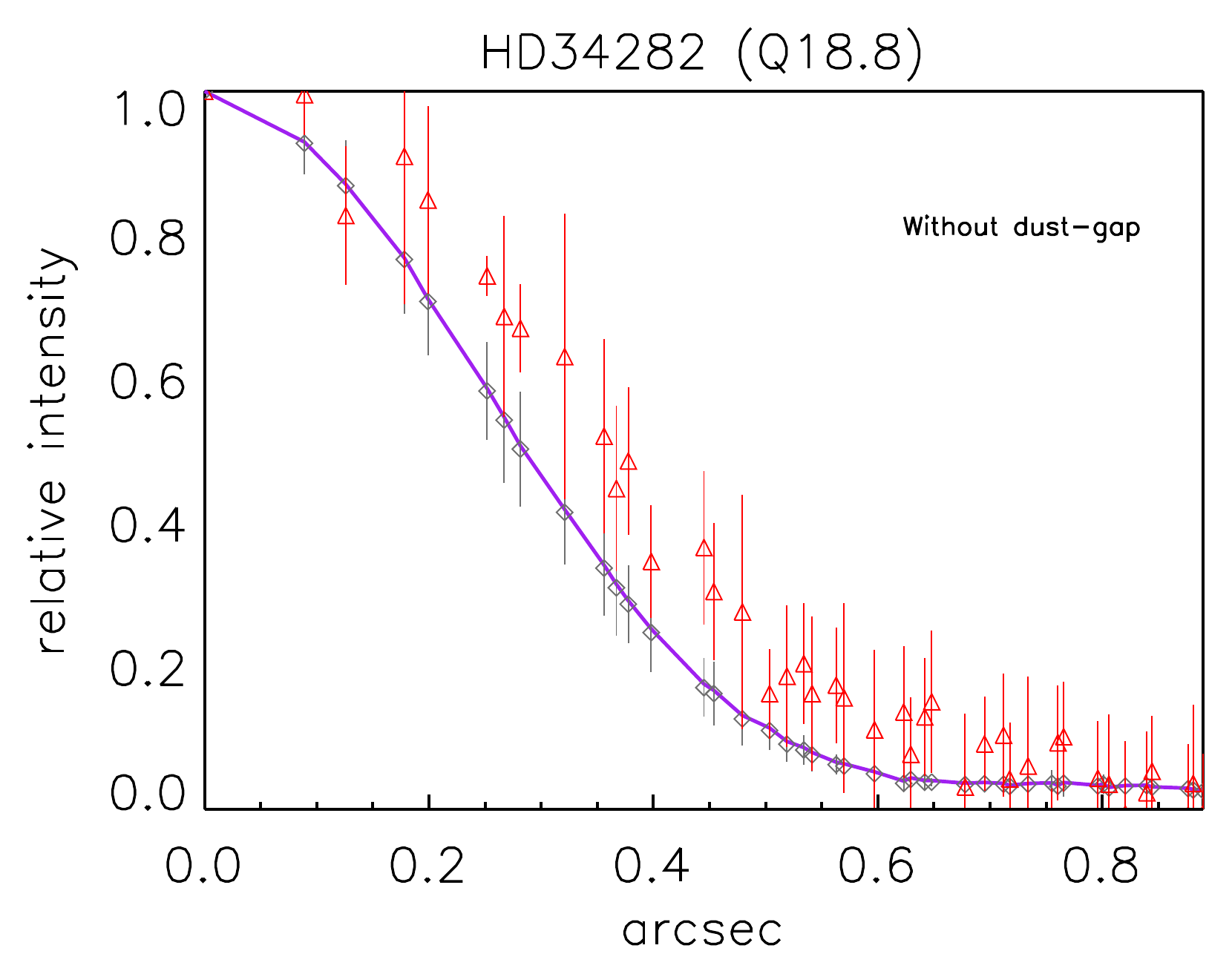}
\includegraphics[width=0.45\textwidth]{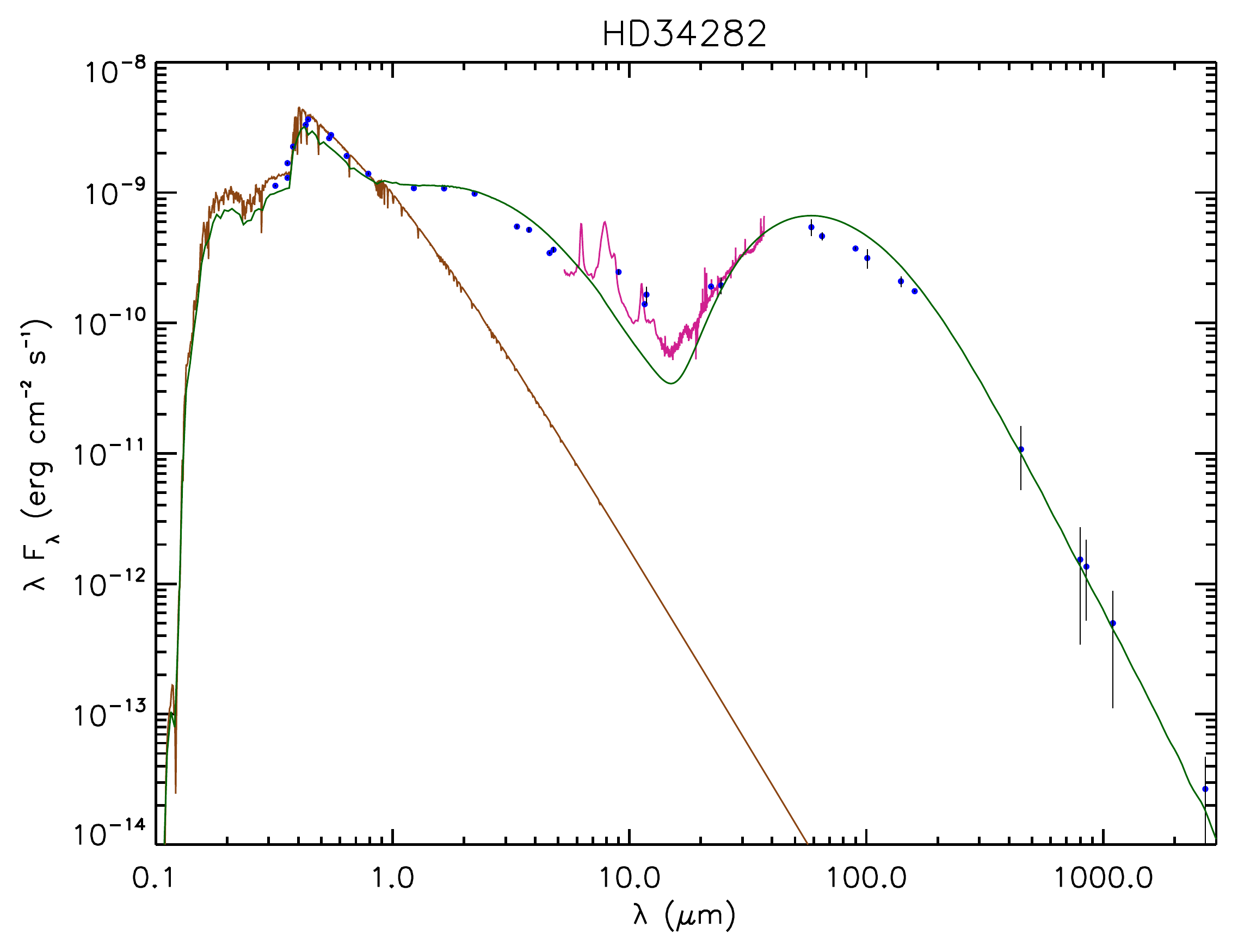}
\includegraphics[width=0.45\textwidth]{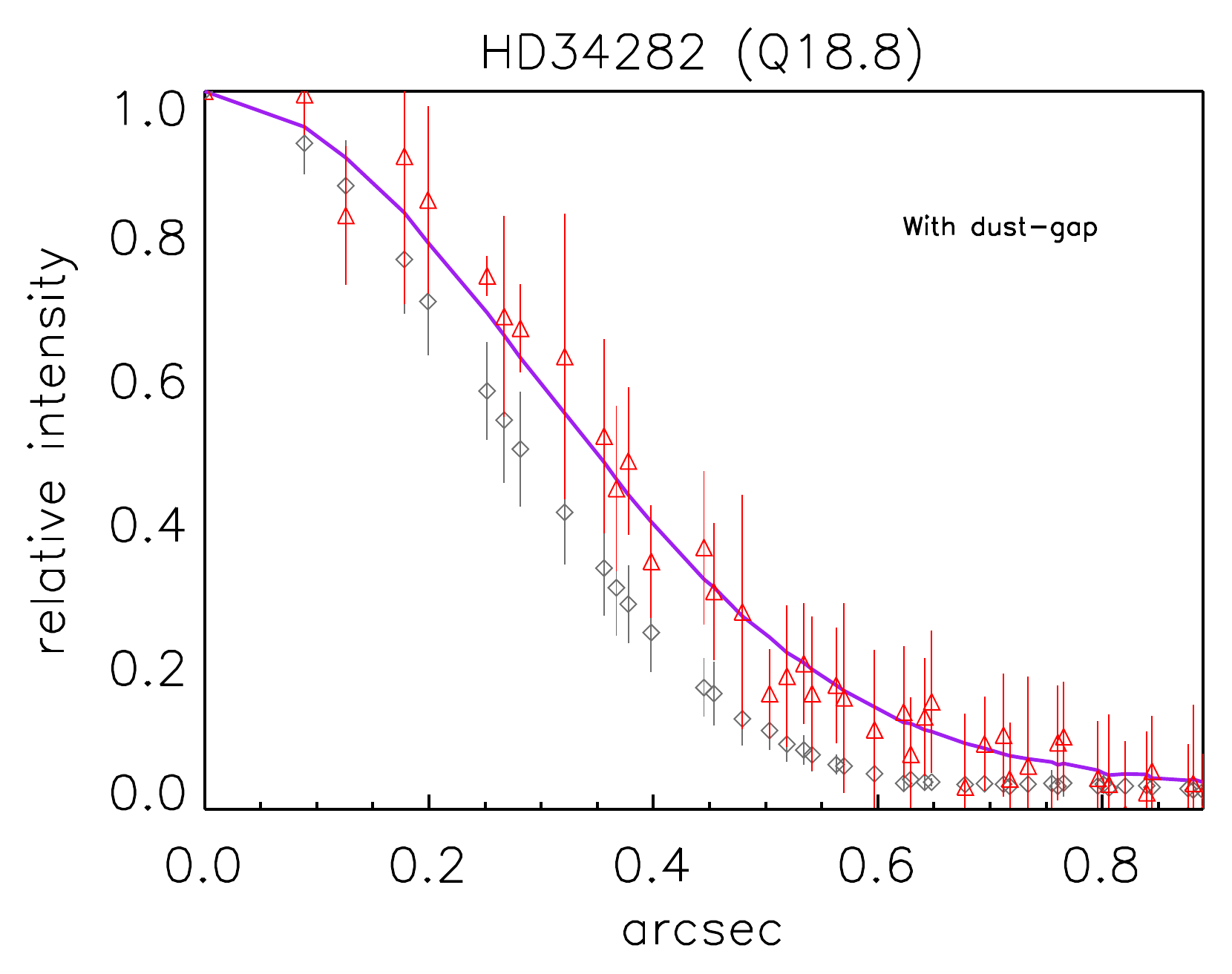}

\caption{\label{models_hd34282} The Spectral Energy Distributions (right) and Radial Brightness Profiles of the Q-band (left) of the continuous disk (top) and transitional disk (bottom) models for HD~34282. Both the SED and the RBP are better fitted by the disk-gap model. See the caption of Figure \ref{models_hd100453} for a description of the lines and symbols shown in this figure. }
\end{figure*}

We collect data of  HD~100453 and HD~34282 which we use for the analysis of this paper. The SEDs and Q-band radial brightness profiles (RBPs) are shown in Figures \ref{models_hd100453} and \ref{models_hd34282}. In addition, we compare these two objects to other similar transitional disks of the sample of Herbig stars presented in \citet{2013Maaskant, 2014Maaskant}.

\begin{table}
\begin{center}
\begin{tabular}{ l l c  c }
\hline
\hline

parameter		&	unit				& HD~100453 	&	HD~34282 \\
		\hline
R.A.		&	(J2000) 			&	11:33:05.58		&	05:16:00.48		\\
Dec.		&	(J2000) 			&      -54:19:28.5 		&	-09:48:35.4		\\
T		&	K				&	7400				&	8625		\\
 L$_{*}$	&	L$_{\bigodot}$ 		&	10.0				&	13.6  \\	
 d		&	pc				&	$114^{+11}_{-9}$	&	$348^{+129}_{-77}$ pc	\\
 M$_{\ast}$&	M$_{\bigodot}$		&	1.66				&	1.59				\\
spectral type & - & A9Ve & A0Ve \\

\end{tabular}
\caption{ \label{tbl-stellar-parameters} Stellar parameters used in this study. The parameters of HD~100453 are taken from \citet{1998Ancker} and \citet{Fujiwara2013}, and the stellar properties of HD~34282 are taken from \citet{2004Merin} and \citet{1998Ancker} }
\end{center}
\end{table}


\subsection{Data}
\label{sec:data}

HD~100453 and HD~34282 were observed as part of a larger high-resolution imaging survey of protoplantary disks around Herbig Ae/Be stars \citep{2011Marinas}. Q-band images (18.1 $\mu$m, $\Delta \lambda=1.1$) were obtained using MICHELLE on the Gemini North telescope on February 2$^{nd}$, 2006 for HD100453 and December 4$^{th}$, 2003 for HD34282. The pixel scale of the observations was $0.089''$. The standard chop-and-nod technique was used to remove thermal background from the sky and the telescope. A point-spread-function (PSF) star, selected from the Positions and Proper Motions (PPM) Catalogue, was observed before and after each science target observation to assess image quality. Mid-infrared standard stars Sirius and HD133774 were used for image calibration and airmass correction. All data was reduced using IDL (Interactive Data Language). 

The observations of the disks around the Herbig stars HD~100453 and HD~34282 are resolved with respect to their point spread functions. For HD100453 the FWHM of the science observation is $0.58 \pm 0.01$ and the FWHM of the PSF is $0.53 \pm 0.02$. For HD34282 the FWHM of the science observation is $0.64 \pm 0.03$ and the FWHM of the PSF =  $0.54 \pm 0.01$. The FWHM values are derived by fitting a Moffat function through the data, where the error is determined by the variability during the observation. For all observational details we refer to \citet{2011Marinas}.

The MIR spectra of the disks are obtained by the Spitzer/IRS telescope and are adopted from \citet{2010Juhasz} and \citet{2010Acke}. The photometric data is taken from the literature and shown in Tables \ref{tab:photometry100} and \ref{tab:photometry342}.  The central star is described by a Kurucz model with the stellar parameters presented in table \ref{tbl-stellar-parameters}. Figures \ref{models_hd100453} and \ref{models_hd34282} show the SEDs and RBPs.

\begin{table}[htdp]
\tiny
\caption{  \label{tab:photometry100} Photometric data HD\,100453  (not corrected for extinction) used in this study }
\begin{center}
\begin{tabular}{l r r @{ $\pm$ } l  c  }
\hline
\hline
band ID  &  \multicolumn{1}{c}{$\lambda$ [$\mu$m]}& \multicolumn{2}{c}{F$_{\nu}$ [Jansky] }& reference \\
\hline

IUE 15  &    0.15 &   $2.00 \times 10^{-3}$ &   0.00 & a \\
IUE 18  &    0.18 &   0.07 &  0.00 & a\\
IUE 22  &    0.22 &   0.17 &   0.00 & a\\
IUE 25  &    0.25 &   0.17 &   0.00 & a\\
IUE 30  &    0.30 &   0.50 &   0.01 & a\\
Johnson U  &    0.36 &   1.08 &   0.03 & b\\
Johnson B  &    0.44 &   2.72 &   0.04 & b\\
Johnson V  &    0.55 &   2.99 &   0.03 & b\\
Near-IR J  &    1.23 &   2.43 &   0.05 & b\\
Near-IR H  &    1.65 &   2.57 &   0.05 & b\\
Near-IR K  &    2.22 &   3.24 &   0.06 & b\\
WISE 1  &    3.35 &   3.87 &   0.30 & c\\
Near-IR L  &    3.77 &   4.77 &   0.23 & b\\
WISE 2  &    4.60 &   7.58 &   0.44 & c\\
Near-IR M  &    4.78 &   4.18 &   0.20 & b\\
AKARI S09  &    9.00 &   5.52 &   0.08 & d\\
WISE 3  &   11.60 &   8.31 &   0.05 & c\\
IRAS 12  &   11.80 &   7.23 &   0.70 & e\\
AKARI S18  &   18.00 &  18.74 &   0.25 & d\\
WISE 4  &   22.10 &  29.13 &   0.19 & c\\
IRAS 25  &   25.00 &  33.53 &   4.09 & e\\
IRAS 60  &   60.00 &  39.38 &  11.35 & e\\
AKARI S65  &   65.00 &  35.23 &   0.96 & d \\
AKARI S90  &   90.00 &  28.65 &   0.81 & d\\
IRAS 100  &  100.00 &  23.82 &   8.31 & e\\
AKARI S140  &  140.00 &  15.77 &   2.12 & d\\
AKARI S160  &  160.00 &  12.81 &   1.22 & d\\
SIMBA 1.2 mm  & 1200.00 &   0.26 &   0.02 & f\\

\hline

\end{tabular}
\end{center}
\textbf{References: }
\textbf{a)}  IUE archival data 
\textbf{b)}   \citealt{1998Malfait}
\textbf{c)} WISE All-Sky Data Release
\textbf{d)} AKARI/IRC mid-IR all-sky Survey 
\textbf{e)} IRAS Point-source catalogue
\textbf{f)} \citealt{2003bMeeus}
\end{table}%

\begin{table}[htdp]
\tiny
\caption{  \label{tab:photometry342} Photometric data HD\,34282  (not corrected for extinction) used in this study }
\begin{center}
\begin{tabular}{l r r @{ $\pm$ } l  c  }
\hline
\hline
band ID  &  \multicolumn{1}{c}{$\lambda$ [$\mu$m]}& \multicolumn{2}{c}{F$_{\nu}$ [Jansky] }& reference \\
\hline
Walraven W  &    0.32 &   0.09 &   0.00 & a\\
Johnson U  &    0.36 &   0.16 &   0.00 & b\\
Walraven U  &    0.36 &   0.12 &   0.00 & a\\
Walraven L  &    0.38 &   0.23 &   0.01 & a\\
Walraven B  &    0.43 &   0.39 &   0.01 & a\\
Johnson B  &    0.44 &   0.44 &   0.01 & b\\
Walraven V  &    0.54 &   0.41 &   0.00 & a\\
Johnson V  &    0.55 &   0.44 &   0.00 & b\\
Cousins R  &    0.64 &   0.36 &   0.01 & a\\
Cousins I  &    0.79 &   0.34 &   0.01 & a\\
Near-IR J  &    1.23 &   0.43 &   0.01 & b\\
Near-IR H  &    1.65 &   0.58 &   0.01 & b\\
Near-IR K  &    2.22 &   0.71 &   0.01 & b\\
WISE 1  &    3.35 &   0.61 &   0.02 & c\\
Near-IR L  &    3.77 &   0.65 &   0.03 & b\\
WISE 2  &    4.60 &   0.52 &   0.01 & c\\
Near-IR M  &    4.78 &   0.58 &   0.03 & b\\
AKARI S09  &    9.00 &   0.74 &   0.04 & d\\
WISE 3  &   11.60 &   0.54 &   0.01 & c\\
IRAS 12  &   11.80 &   0.65 &   0.1 & e\\
WISE 4  &   22.10 &   1.40 &   0.02 & c\\
IRAS 25  &   25.00 &   1.58 &   0.23 & e\\
IRAS 60  &   60.00 &  10.60 &   1.57 &e\\
AKARI S65  &   65.00 &  10.04 &   0.73 & d \\
AKARI S90  &   90.00 &  11.17 &   0.26 & d\\
IRAS 100  &  100.00 &  10.58 &   1.85 & e\\
AKARI S140  &  140.00 &   9.75 &   0.93 & d\\
AKARI S160  &  160.00 &   9.35 &   0.19 & d\\
450 micron  &  450.00 &   1.61 &   0.83 &b \\
800 micron  &  800.00 &   0.41 &   0.32 & b\\
1100 micron  & 1100.00 &   0.18 &   0.14 & b\\
PDB 1.3 & 1300.00 & 0.10 & 0.02 & g\\
2600 micron  & 2600.0 &   0.02 &   0.019 & f \\
PDB 3.2 & 3200.00 & 0.01 & 0.00 & g\\
\hline

\end{tabular}
\end{center}

\textbf{References: }
\textbf{a)}   \citealt{2001deWinter}
\textbf{b)}   \citealt{1996Sylvester}
\textbf{c)} WISE All-Sky Data Release
\textbf{d)} AKARI/IRC mid-IR all-sky Survey 
\textbf{e)} IRAS Point-source catalogue
\textbf{f)} \citealt{2000Mannings}
\textbf{g)} \citealt{2004Natta} 


\end{table}%

\subsection{The SEDs of transitional disks }
The mid-IR parts of the SEDs of group Ib Herbig Ae/Be stars show similar shapes (see Figure \ref{fig:all}).  All objects show an emission bump at $\sim$20 $\mu$m and PAH features. In HD~100453, a weak signature of the amorphous 10 and 20 $\mu$m  silicate features can be seen (Figure \ref{fig:100453zoom}). All other sources show no sign of amorphous silicate features. In the next section we will characterise the dust gaps in the disks of HD~100453 and HD~34282 and confirm that the shape of the silicate feature is connected to the presence of large dust gaps. The detailed shape of the SED is degenerate as it depends on parameters such as density structure and dust composition and grain size distribution. Though, the `bump' in the the SED at MIR wavelengths for these objects is an indicator of large dust gaps in the disks of group Ib Herbig Ae/Be stars. Radiative transfer modeling of the Q-band is needed to constrain the radii of the inner edges of the outer disks of HD~100453 and HD~34282.


\section{Modeling}
\label{sec:modeling}
In this section we introduce the radiative transfer code MCMax and the dust model. We discuss the modeling approach and outline the parameters that we study in this paper.

\subsection{Radiative transfer code MCMax }
For the modelling of the disks, we use MCMax dust modelling and radiative transfer tool \citep{2009Min}. MCMax performs radiative transfer using a Mont Carlo recipe outlined by \citet{2001BjorkmanWood}. It solves the temperature structure and density structure assuming a 2-D geometry in radial and vertical directions. It is used for modelling of circumstellar material including high optical depth regions and axi-symmetry is assumed for the dust disk model. Our aim is to fit models to the observed Q-band images and SEDs of our sample stars to derive the disk structure. Therefore, we azimuthally average the Q-band brightness profiles and evaluate the radial extent of the disk emission. Generally, the input parameters of the code can be divided in five categories: the stellar parameter, the surface density setup, the opacity parameters and the parameters of the halo, inner disk and the outer disk. The inner wall of the outer disk is puffed up since the high vertical surface brightness causes the temperature in the wall to be higher. This effect is taken into account in the radiative transfer code, which solves for the temperature and density in the disk and interpolates between the chosen grid points of the inner and outer disk. The most important disk parameters that we adjust to fit the model to the observations are: the inner and outer radius of the halo, the inner radius of the outer disk (wall radius), the outer radius of the disk, the mass of the dust in the disk and halo, the grain properties and the power law indices of the opacity profile.
An extensive description of the modeling approach can be found in \citet{2013Maaskant}.

\subsection{Dust model}
The composition of the grains in the disk are 20 \% carbon and 80 \% silicates. The standard dust composition with reference to the optical constants, is 32\% MgSiO$_3$ \citep{1995Dorschner}, 34\% Mg$_2$SiO$_4$ \citep{1996HenningStognienko}, 12\% MgFeSiO$_4$ \citep{1995Dorschner}, 2\% NaAlSi$_2$O$_6$ \citep{1998Mutschke}, 20\% C \citep{1993Preibisch}. The shape of our particles is irregular and approximated using a distribution of hollow spheres (DHS, \citealt{2005Min}) using a vacuum fraction of 0.7. We performed some test modeling, changing the compositional (carbon and silicates) abundances. Although the SED may be sensitive to this parameter, the Radial Brightness Profiles (i.e. images) of these models give comparable dust gap sizes. As a result, the dust composition fractions does not play a role in the derivation of the dust gap size (See Appendix \ref{sec:appendix}).

 \subsection{Model procedure}
 To fit the SED as well as the Q-band size we follow the fitting procedure as outlined in \citet{2013Maaskant}. We summarize the procedure briefly here. As a first step, we start with a disk which has a continuous density profile. We assume that the disk is in hydrostatic equilibrium and that the radial dependence of the dust surface density drops off proportional to a powerlaw of $-1$. We fit the far-Infrared (FIR) to mm photometry to a grain size powerlaw index of p between 3.0 and 4.0. If this does not fit the Q-band size, than we insert a dust gap in the disk. We decrease the surface density in the dust gap by about 3 orders of magnitude. Recall, that previous studies by \citet{2013Maaskant} show that the observation require a contrast of 2 orders of magnitude. This will result in a ``wall'' structure at the inner edge of the outer disk. Now we choose the radius of the inner edge of the outer disk, so that the convolved model image fits the observed Q-band image size. Then, we tried to fit the emission in the NIR by including an optically thick inner disk. A hydrostatic inner disk does not give enough flux to fit the SED. We tried to fit the NIR flux by parameterizing and 'puffing up' the inner disk (i.e. increasing the vertical scale-height and testing several density slopes). We have modeled this scenario extensively and showed the SEDs of these models in Figures A1 in Appendix A. However, we consistently fail to get a good fit for the outer disk because a higher inner disk casts a shadow on the outer disk and therefore reduces the flux at MIR and FIR wavelengths. A parameterized inner disk may fit the NIR flux, however it is inconsistent with an outer disk which is in hydrostatic equilibrium because the vertical scale height of the outer disk is not high enough to receive enough radiation from the central star. To solve for this problem we replace the inner disk with an optically thin inner spherical halo to fit the NIR flux. We recognize that the adopted geometry is a choice of con- venience and many spatial distributions will produce similar fits as long as the structure is optical thin. Conversely, as the dust structure in the inner region has to be optically thin, the particular choice has no influence on our derived quantities such as the size of the dust gap. As a final step, we choose the minimum size (between 0.1 $\mu$m and 1 $\mu$m) of the grains in the disk to fit the flux in the MIR and FIR to the SED.


\section{Results}
\label{sec:results}
Our main focus is to demonstrate the existence of disk dust gaps. The second objective is to derive the location (i.e. size) of the gap. In this section we constrain the disk gap and fit the size of the extended emission in the resolved Q-band images. We present the best fitting radiative transfer models to the Q-band sizes and the SEDs of HD~100453 and HD~34282.

\subsection{Constraining the dust gap size}

It has been shown by \citet{2013Maaskant} that the disk parameter that is constrained by fitting the Q-band size is the inner radius of the outer disk (i.e. the location of the wall). In Appendix \ref{sec:appendix}, we perform a parameter exploration around the best-fit solution to study the effect on the SED of the inner and outer radii of the outer disk, the mass of the outer disk and the grain properties. We confirm that only the inner radius of the outer disk substantially effects the Q-band size. In the remainder of this section we fit the Q-band and SED of HD~100453 and HD~34282. Note that the 1 sigma error bars on the RBPs of Figures \ref{models_hd100453} and \ref{models_hd34282} are a representation of the asymmetry of the image. The actual fitting of the image has been done by a least square calculation to all the pixels, where the photometric error on each pixel is 10\%. We fit a Moffat function through the Q-band image of our model and make sure that the model FWHMs are consistent with the FWHMs of the observations. The typical errors of these fits are in the order of a few percent.

\begin{table*}[t]
\begin{center}
\begin{tabular}{l  c c c c c c c }
\hline
\hline
Object & M$_{dust}$ & M$_{halo}$& R$_{inner~disk/halo}$ & R$_{wall}$ & R$_{out}$ & a$_{min}$~--~a$_{max}$ & a$_{pow}$ \\
&  M$_{\odot}$&  M$_{\odot}$ & AU & AU & AU &  &  \\
\hline
HD100453 & 3.2$\times$10$^{-4}$  & 0.15$\times$10$^{-9}$  & 0.1 $-$ 1.7 & $20^{+2}_{-2}$ & 200 & 0.5 $\mu$m $-$ 1mm  & -3.5\\
HD 34282 & 6.1$\times$10$^{-4}$& 5.5$\times$10$^{-11}$  & 0.05 $-$ 1.3 & $92^{+31} _{-17}$ & 900 & 0.5 $\mu$m $-$ 1mm & -3.5\\
\hline
HD\,97048	&	6.0$\times$10$^{-4}$	&	\dots	&	0.3 $-$ 2.5			&	34$^{+4}_{-4}$ 	& 500				& 0.5 $\mu$m $-$ 1mm	& -3.5\\	
HD\,169142	&	0.8$\times$10$^{-4}$	&	0.31$\times$10$^{-12}$	&	0.1 $-$ 0.2			&	23$^{+4}_{-4}$		& 235 				& 0.5 $\mu$m $-$ 1mm	& -3.5	\\	
HD\,135344\,B 	&	1.0$\times$10$^{-4}$	&	0.47$\times$10$^{-12}$	&	0.1 $-$ 0.3 			&	30$^{+4}_{-3} $ 	& 200				& 1.0 $\mu$m $-$ 1mm	& -4.0	\\	
Oph IRS 48	&	3.0$\times$10$^{-5}$	&	0.50$\times$10$^{-12}$	&	0.1 $-$ 0.3 			&	63$^{+4}_{-4} $	& 235				& 0.1 $\mu$m $-$ 1mm	& -4.0		\\	

\end{tabular}
\caption{ \label{tbl-best-fit-model-parameters} Best-fit model parameters and comparison to the previous values by \citealt{2013Maaskant}} 
\end{center}
\end{table*}

\subsection{Best-fit model HD~100453}

We first explore a continuous disk model (i.e. without a gap in the disk). This seems to fit the SED reasonably well, though it always fails to fit the size of the Q-band image (see Figure \ref{models_hd100453} top panels). We find that this is because for a continuous disk, the disk surface which is emitting in the Q-band is much closer to the star (i.e. $\lesssim 10$ AU). Therefore, the model Q-band image is barely resolved with respect to the PSF. 

\begin{figure}
\includegraphics[scale=0.5]{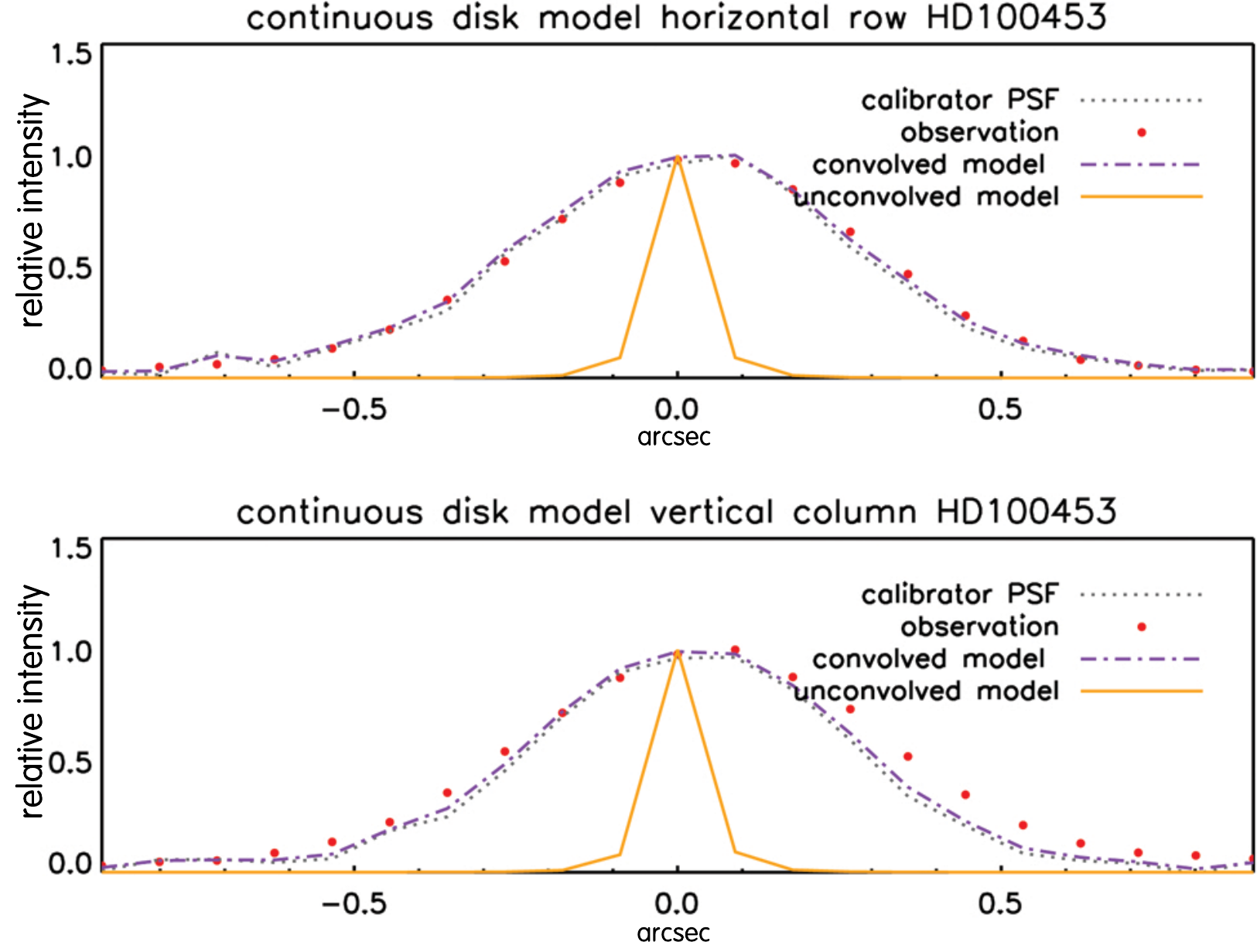}
\includegraphics[scale=0.5]{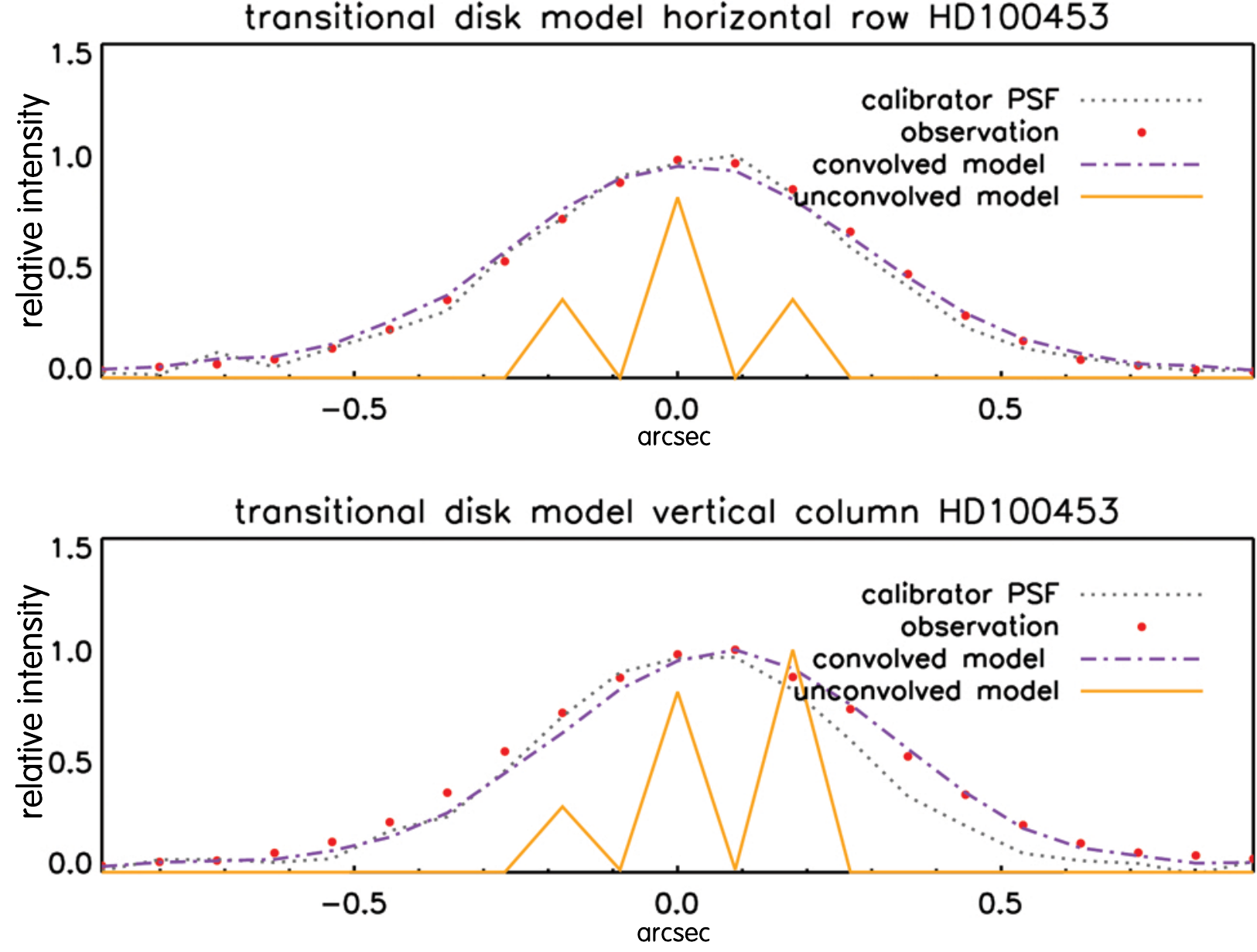}
\caption{HD~100453: horizontal and vertical radial brightness profiles of the continuous (top) and transitional disk models (bottom).}
\label{RBP100453-withgap-nogap}
\end{figure}

\begin{figure}
\includegraphics[scale=0.4]{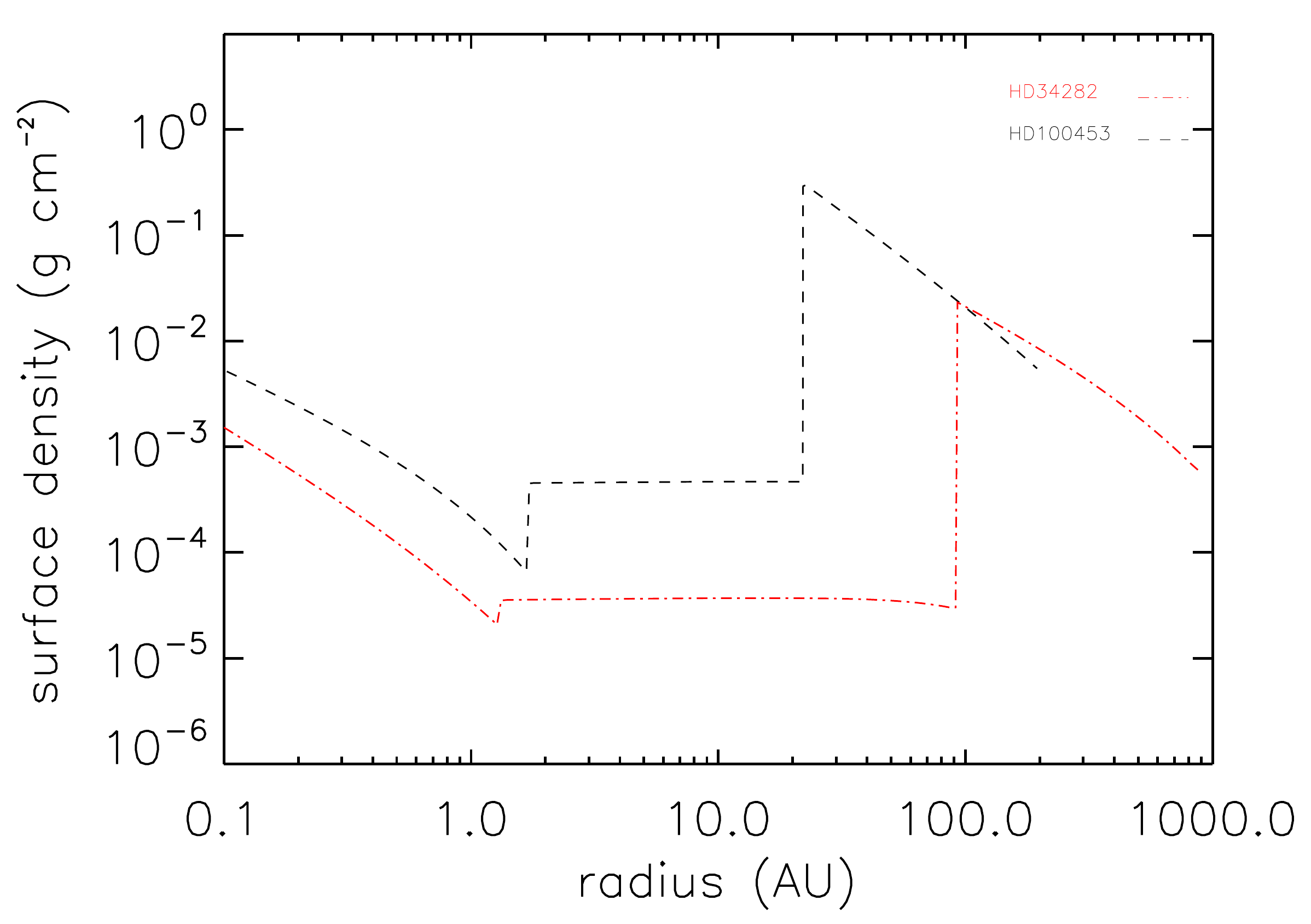}
\caption{Surface density as a function of radius for our best fitting model including an inner halo, a gap and a hydrostatic outer disk.}
\label{fig:surface_density}
\end{figure}

We can only fit the size of the Q-band by including a dust gap in the disk. The model that gives us the best-fitted SED  and radial brightness profile (Figure \ref{models_hd100453}-bottom) has a large dust gap in the disk. We find that the inner edge of the outer disk has a very high surface brightness and dominates the emission in the Q-band. For this reason, the Q-band size is very sensitive for the location of the inner edge of the outer disk. We constrain its location at $20^{+3}_{-3}$ AU. The surface density of our the best fitting model is shown in Figure \ref{fig:surface_density}. The inclination of HD~100453 is not known in the literature. This introduces the dominant uncertainty in the location of the inner edge of the outer disk. In our simulation we have assumed an inclination of 45 degrees. For a pole-on or (nearly) edge-on geometry, we find that the location of the wall must be respectively decreased or increased by 3 AU to fit the size of the Q-band again. As can be seen in Figure \ref{models_hd100453}-bottom, the convolved image that is modelled using these input parameters fits well with this observation. The average temperature of the inner edge ($20 - 23$ AU) of the outer disk is $\sim160\pm10$ K.

In Figures \ref{RBP100453-withgap-nogap}, horizontal and vertical radial brightness profiles of the continuous disk (top) and transitional disk (bottom) models are compared to the observation. In the horizontal direction both convolved models, with and without gap equally agree with the observation. However, in the vertical direction, the convolved image of the transitional disk model provides a significantly better fit. One would expect for inclined transitional disks that the farther side is brighter as observed from Earth. The asymmetry is evidence that the disks are inclined, where we recognise that deriving an inclination angle will be hampered by the uncertainty in the vertical height of the disk and is beyond the scope of this paper. The transitional disk model slightly underestimates the left side of the RBP in the vertical direction. This asymmetry could be indicative of extra emission associated with additional structure in the disk that is not considered in the model. Further observations are required to clarify this issue.

The structure of the inner disk is poorly constrained by our data. Therefore we have modelled the inner circumstellar region by spherical halos which are optically thin, but geometrically high. To fit the near-infrared (NIR) emission we adopt a halo from 0.1 to 1.7 AU with a mass of $0.15\times10^{-9}$ M$_{\odot}$. As there are many uncertainties in the structure of the inner disk (e.g. \citealt{2010DullemondMonnier}), we only use the halo in our model in order to reproduce the amount of produced NIR emission from the disk. To account for the absence of silicate features at 10 and 20 $\mu$m, we assume a dust species in the halo which does not show any features. We adopt 100\% amorphous carbon, but alternatively, larger silicate grains or metallic iron may yield similar results. 
Unfortunately no interferometric dataset exists yet for HD34282  and HD100453. In our modeling process we have tested the possibility of 100\% carbon disks closer to the star. However, hydrostatic inner disk models (or even vertically puffed up by a factor of 2-3) do not fit the near-infrared SED. Actually our `halo' solution drops the assumption of hydrostatic and is therefore similar to previous studies which have shown to fit the NIR part of the SED by including an optically thin inner disk \citep{2011Verhoeff, 2011Mulders, 2013Maaskant}.

Clearly, the absence of small silicate grains in the inner region of protoplanetary disks is remarkable. Similar dust compositional solutions have been inferred for a number of other transitional disks (e.g. T Cha: \citealt{2013Olofsson}, HD\,135344\,B: \citealt{2014Carmona}), this could be because close to the star, only refractory grains are present because small silicate grains have been sublimated. So not only the correlation with the presence of dust gaps, but also the assumption that there are no small silicate grains in the optically thin halo, causes the absence of silicate features in the SED.

The values of the best-fit parameters are listed in Table \ref{tbl-best-fit-model-parameters} and also compared with the values obtained for four other Herbig star by \citet{2013Maaskant}.

\subsection{Best-fit model HD~34282}

For the the Herbig star HD~34282, we follow the same procedure as HD~100453. We find that we can not find a fit to the SED and Q-band size by using a disk with a continuous density structure (Figure \ref{models_hd34282}-top panels).

To fit the SED and RBP, we adopt a dust gap in the disk. The best fit to the SED and RBP can be seen in Figure \ref{models_hd34282}-bottom panels. The location of the inner radius of the outer disk is constrained to $92^{+31} _{-17}$ AU. Although the signal to noise of the Q-band observation is worse than that of HD~100453, the error is still dominated by the relatively large error in the uncertainty of the distance.

\begin{figure}
\includegraphics[scale=0.5]{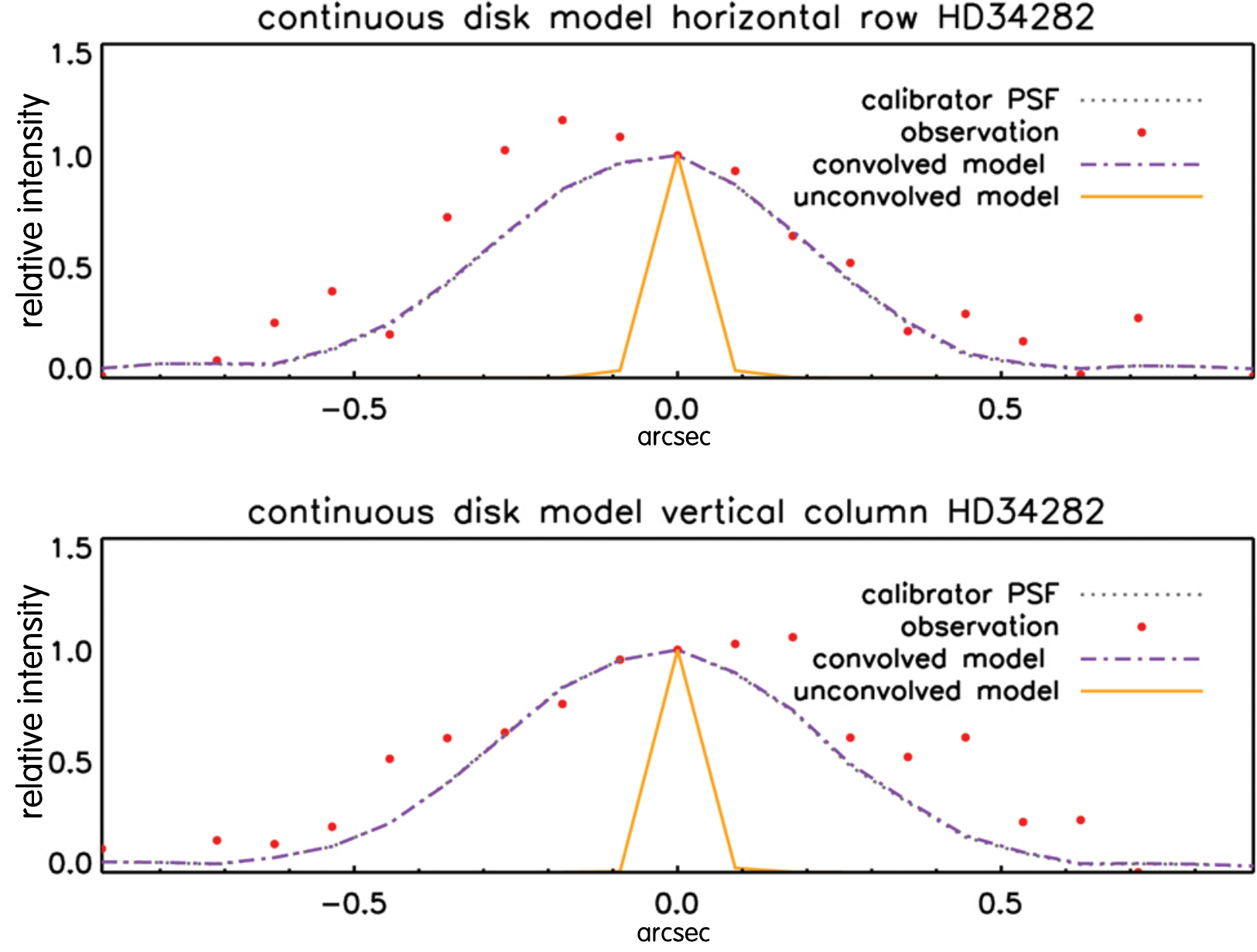}
\includegraphics[scale=0.5]{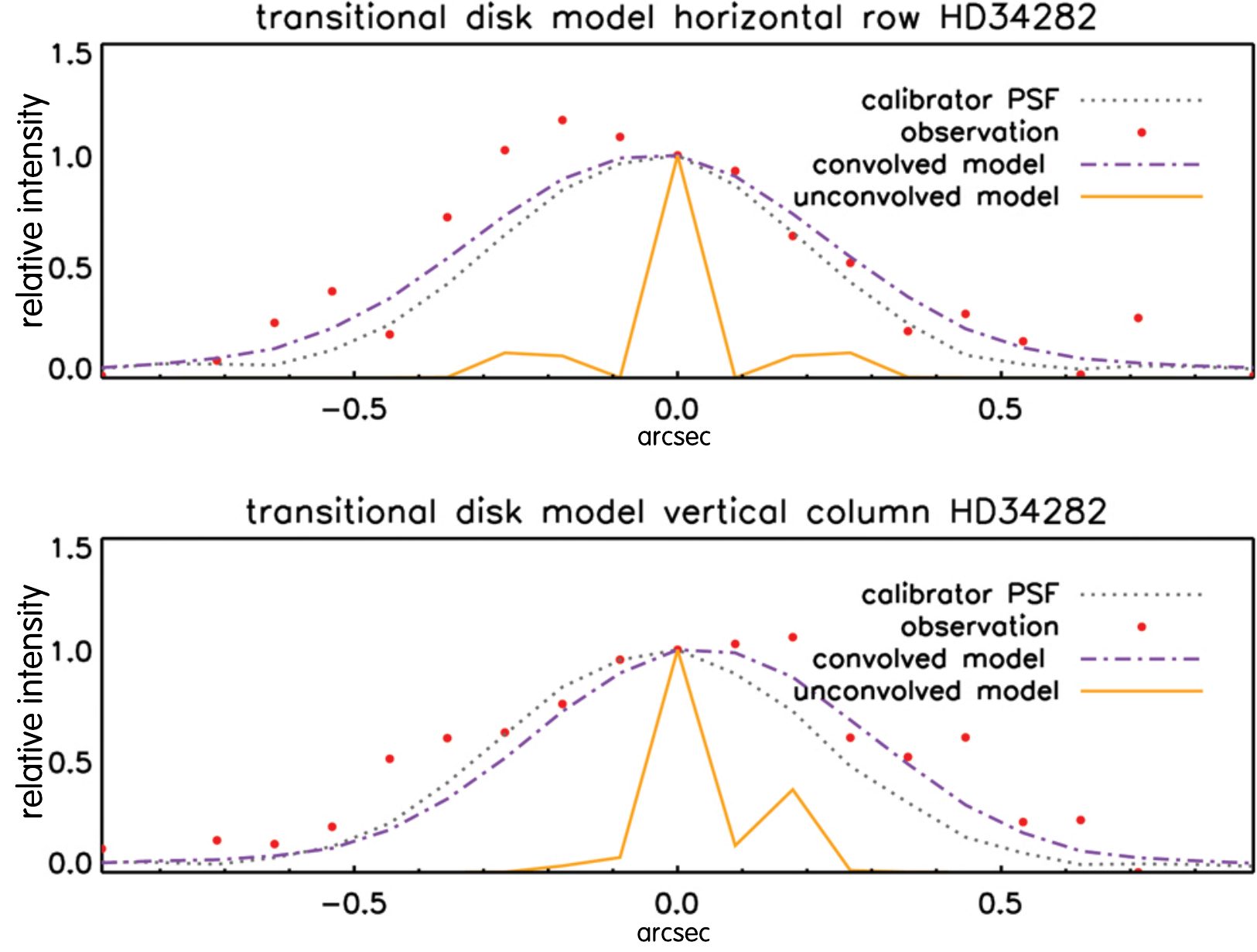}
\caption{HD~34282:  horizontal and vertical radial brightness profiles of the continuous (top) and transitional disk models (bottom).}
\label{RBP34282-withgap-nogap}
\end{figure}

The RBPs in the horizontal and vertical directions of the continuous (top) and transitional (bottom) disk models are compared in Figure \ref{RBP34282-withgap-nogap}. The scatter in the image for this target is high and therefore we have azimuthally averaged the RBPs in all directions to fit the best model (shown in Figure \ref{models_hd34282}). It is still evident however, that the observation is resolved in the horizontal and vertical directions. Therefore, the model with gap agrees best to the observation compared to the model without gap.

The model of the disk includes a halo that is extended from 0.02 to 1.3 AU and has a mass of $2.4\times 10^{-11}$ M$_{\odot}$. The dust grains in the halo are made of carbon and they have the size of 0.5 $\mu$m to 1 mm with a size distribution power law of -3.5. The mass of the outer disk is considered to be $0.83\times 10^{-4}$ M$_{\odot}$. A model with an inner disk instead of halo does not fit well, since such solution does not provide enough flux to fit the NIR SED. However with the halo model the flux level can be increased sufficiently. The surface density of the best fitting model is shown in Figure \ref{fig:surface_density}.

Alternatively we could also model a hydrostatic inner disk, though that model would require the outer disk to be puffed up out of hydrostatic equilibrium. We choose an optically thin but geometrically high solution for the inner region (Honda 2012). However, if we drop the assumption of hydrostatic equilibrium and simply parameterize the \textit{entire} disk, we can model an inner disk as well. A parameterized inner disk requires that the outer disk is also parameterized, because the vertical scale height of the inner and outer disk needs to be several times higher than prescribed by hydrostatic equilibrium to fit the SED. But we discard this model because we do not want to loose the assumption of hydrostatic equilibrium. A parameterized model would require a puffed up inner disk and wall.  It should be mentioned that this discussion applies to both objects. 

\subsection{Upper limit on the dust mass in the gap}
\label{sec:upper limit}

In agreement with previous studies \citep{2013Maaskant}, for our best fit to the SED and Q-band image, we have derived that the minimum difference in surface density between the dust gap and the outer disk is roughly three orders of magnitude. We examined the  maximum dust mass distribution that can be hidden in the gap without influencing our results. We find that if we increase the dust mass in the gap above $\sim10^{-7}$M$_{\odot}$ for both sources, the observed size of the image in the Q-band cannot be reproduced anymore. We have chosen a constant density distribution as a function of radius for the dust in the gap to prevent the dust to become optically thick. This reflects the increased emission from the inner region. The surface density distribution is shown in Figure \ref{fig:surface_density}. In addition - and actually the dominant effect - since the inner region becomes optically thick, the stellar photons do not reach the wall anymore and the height of the wall decreases. This also results in a decreased Q-band size. This effect is also quite obvious in the SED and for these gap masses the SED fits also deviate strongly from the observations in the Q-band region (see Figures \ref{upper limit_HD100 appendix} and \ref{upper limit_HD342 appendix} and appendix \ref{upper appendix}). Future observations with ALMA should be able to observe the contrast between the dust gap surface density and the outer disk.

\subsection{Source asymmetry}

It is not our aim to fit the source asymmetry or the influence of the structure of the wall on the observation because the MIR observations do not have sufficient resolution to fit such sub-structures. In our model, a sharp vertical inner radius of the outer disk is assumed. This vertical assumption may not be completely realistic. We have tested this by varying the wall shape in MCMax and, at the resolution of our data, we see little variation in the Q-band images. We suggest that interferometric observations may be able to track the structure of the inner wall of the outer disk.

\subsection{Summary}

By using radiative transfer models, we have studied the Q-band sizes and SEDs of HD~100453 and HD~34282. We fail to find solutions with a continuous density structure. Thus we find that it is required to include large dust gaps in the disk to simultaneously fit the Q-band size and SED. The inner edge of the outer disks of  HD~100453 and HD~34282 are respectively $20^{+3}_{-3}$ AU and $92^{+31} _{-17}$ AU.  

\section{Discussion}
\label{sec:discussion}

For HD~100453 and HD~34282, radii of the inner edges of the outer disks have been derived from the best-fit models. These values are in the same range as the wall radii of disks around other group Ib Herbig stars. We now discuss the implications of these results.

\subsection{New Meeus group classification based on the $F_{30}/F_{13.5}$ continuum flux ratio}

\begin{figure}[t]
\includegraphics[scale=0.6]{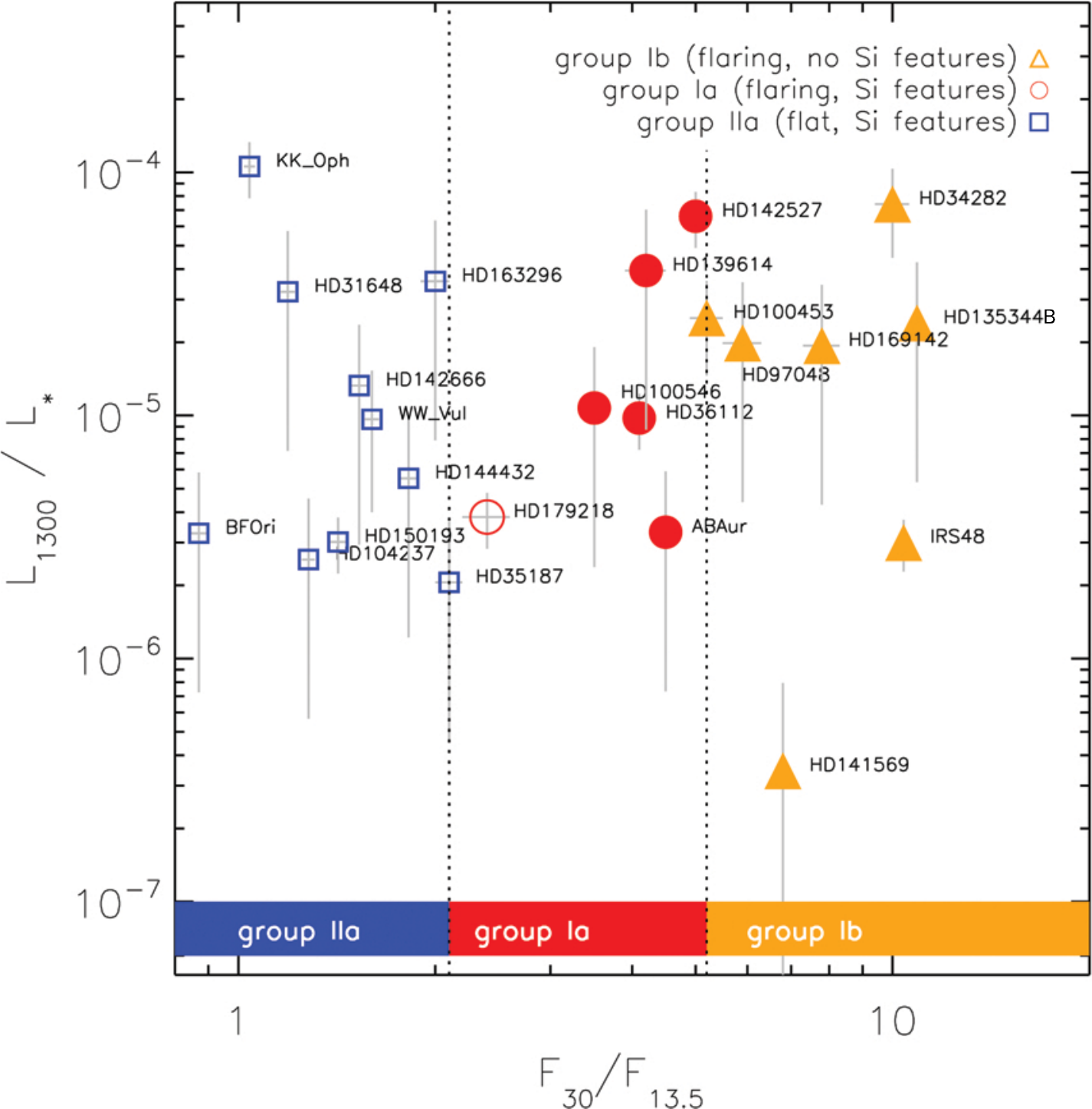}
\caption{\label{fig:color-mag} The luminosities at 1.3 mm is compared to the flux ratio at 30.0 and 13.5 $\mu$m. The filled symbols indicate whether large dust gaps have been detected in the literature. The large dust gaps are deduced from both direct sub-millimetre imaging and indirect method of SED and mid-IR image fitting suggested in this paper. The $L_{1300}/L_{*}$ can be used as a proxy of the disk mass. We confirm the connection between the dust gap size and its location, with the $F_{30}/F_{13.5}$ ratio. The groups are divided by $2.1 < F_{30}/F_{13.5} < 5.1 $. All the objects in this diagram are taken from the sample of \citet{2014Maaskant}.  }
\end{figure}

We compare the photometric properties of HD~100453 and HD~34282 to other Herbig stars. Figure \ref{fig:color-mag} shows the color-magnitude and classification of a sample of Herbig stars. It shows luminosities at 1.3 mm compared to flux ratios at 30.0 and 13.5 $\mu$m for Herbig stars from the sample of \citet{2014Maaskant}.  As disks lose their mass during their evolution, they move down in this diagram. This diagram seems to indicate that there is no trend between group I and group II disks with regard to the disk mass.  Previous studies \citep{2009Acke, 2013Maaskant, 2014Maaskant} have indicated a strong correlation between the $F_{30}/F_{13.5}$ continuum flux ratio and the group classification based on \citep{2001Meeus}. The transitions between group Ia, Ib and group IIa may be more easily expressed by the the $F_{30}/F_{13.5}$ ratios. For flat sources with amorphous silicate features (group IIa) that implies $F_{30}/F_{13.5} \leqslant 2.1$. For transitional/flaring objects with amorphous silicate features (group Ia) we find $2.1 < F_{30}/F_{13.5} < 5.1 $. The transitional/flaring objects which do not have silicate feature (group Ib) have $F_{30}/F_{13.5} \geqslant 5.1$. The Meeus groups are well separated by the new criteria based on the mid-infrared flux ratio \ref{fig:color-mag}.

\subsection{Correlation between amorphous silicate emission, size of the dust gap and the $F_{30}/F_{13.5}$ ratio}

Transitional flaring disks which do show silicate emission are group Ia Herbig Ae/Be stars. The amorphous silicate features in these objects may originate in the inner edge of the outer disk. This may be the case for HD~100456 (e.g. \citealt{2011Mulders}) and HD~139614 \citep{2014Matter}, two transitional objects with the inner edge of the outer disk at respectively 13 and 5.6 AU. Alternatively, the amorphous silicate feature may originate in the inner disk. Examples are HD~142527 \citep{2011Verhoeff}, AB Aur \citep{2010Honda} and HD~36112 \citep{2010Isella}. Also in these objects large dust gaps of several tens of AUs have been found. The large dust gaps are deduced from both direct sub-millimetre imaging and indirect method of SED and mid-IR image fitting suggested in this paper. However, the inner disk may still contain a substantial amounts of small silicate grains producing the silicate features. The following objects are previously resolved using sub-mm imaging: HD~135344~B \citep{2009Brown}, HD~142527 \citep{2006Fukagawa} and IRS~48 \citep{2013vanderMarel}.

Group Ib Herbig Ae/Be stars do not show silicate emission features. As figure \ref{fig:color-mag} shows, the weakness of silicate emission features is connected to the presence of large disk dust gaps in the critical temperature regime ($\gtrsim$160 K) responsible for emission of silicate features. HD~100453 has the smallest dust gap size in the sample of group Ib objects. Peculiarly, a closer look at the Spitzer/IRS spectrum of HD~100453 may show a tentative detection of  very weak amorphous silicate features (Figure \ref{fig:100453zoom}). Figure \ref{fig:color-mag} shows that the $F_{30}/F_{13.5}$ ratio of HD100453 is on the border of a group Ia identification. We suggest that this is connected to having smaller dust gap compared to the other disks and thus higher temperature in the inner region of the outer disk. Our radiative transfer model of HD~100453 shows that the average temperature of the inner region of the outer disk (between $20 - 23$ AU) is $\sim$160 K. Because the inner edge of the outer disk is closer to the star, the temperature of the dust in the wall is higher than for HD~34282 and the other group Ib transitional disks studied in \citet{2013Maaskant}. We can compare the temperature in the wall of HD~100453 to that of HD~100546 and HD~97048. HD~100546 is one of the best studied transitional disks which does show strong amorphous silicate features. Radiative transfer models presented in \citet{2011Mulders} show that the inner edge of the outer disk of  HD~100546 is located at $\sim$13 AU and has a temperature of $\sim$200 K. For HD~97048, the typical temperature in the inner region of the outer disk ($34-37$ AU) is $\sim$110 K \citep{2013Maaskant}. HD~97048 does not show any sign of amorphous silicate emission. From this comparison, we infer that the temperature transition from $\sim$160 K to $\sim$200 K in the inner edge of the outer disk is critical for the strong enhancement of the amorphous silicate features originating in the wall. Possibly this is connected to the fact that grains of $\lesssim$160 K are icy and therefore grow to larger typical sizes (e.g. \citealt{2011Sirono,2012Okuzumi}). Models predict that across the snow-line, the cycle of sublimation and condensation will allow efficient growth and trapping (e.g. \citealt{2013RosJohansen}). The slow sublimation of ice in icy aggregates at $\sim160$ K, which are radially drifting inward in the disk, will lead to the formation of ``pure'' silicate aggregates which are thus more fragile and can readily fragment upon collision. In this scenario, an enhanced abundance of small fragmented silicate grains at temperatures of $\gtrsim160$ K, may contribute significantly to the amorphous silicate features. The role of this particular snowline in the evolution of solids may be further investigated by direct observations of the snowline such as for HD16296 \citep{2013Mathews} and TW Hya \citep{2013Qi}.

\subsection{Evolutionary link between transitional (flaring) and self-shadowed (flat) disks.}

A key question in the study of protoplanetary disks is to determine the evolutionary link between dust gaps formation (group I) and grain growth and settling (group II). Disk dust gaps are found in an increasing number of group I Herbig stars and there is yet no evidence of large dust gaps in group II Herbig stars \citep{2013Maaskant}. Therefore, we speculate that it is very unlikely that group I evolves to group~II. There are now two possible evolutionary scenario's to understand the link between group I and II. First, as proposed by \citet{2013Maaskant}, both groups may have evolved from a common ancestor (i.e., gapless flaring-disk structure). In transitional group I objects, dust gap formation has preceded the collapse of the outer disk while grain growth and dust settling have flattened the outer disk in flat group II objects. Secondly, group II objects may be the precursors of group I objects. In that case, possibly planet formation is followed by the formation of a large dust gap that may produce a high vertical wall and stir up the dust in the outer disk.
In order to distinguish the proposed observational scenarios, future observational at scattered light wavelengths up to the (sub-)mm are required to search for dust depletion and the interaction with gas flows through the gap. If, for example, circumplanetary-disks around newly forming planets will be found than this enable us to understand the link between planet formation and proto-planetary disk evolution. Another step forward in distinguishing between these evolutionary scenarios may be expected from observations of the outer disks of group II objects. Yet so far, their structure is completely unknown as no observation has been sensitive to detect the outer disks and characterise their typical radial size. If these disks are radially small (~10 AU) but very dense and optically thick, than these objects must have had a different evolution as compared to radially large group I objects. However, if these objects are radially large as well, then they may still be precursors to group I objects. ALMA is needed to understand if these objects are `small and fat' or `large and flat.'


\section{Conclusions}
\label{sec:conclusions}
We have used spatially resolved MIR observations and radiative transfer models. We fit the spectral energy distribution and the radial brightness profile of the Q-band images of the disks around two Herbig Ae/Be type Ib stars, HD~100453 and HD~34282. This work is compared to the results of \citet{2013Maaskant}, where a similar analysis of four other Herbig stars, HD~97047, HD~169142, HD~135344 B and Oph~IRS~48, is presented.

\begin{itemize}

\item We were not able to fit the spatially resolved Q-band imaging and  SEDs considering a continuous disk (i.e. no gap). In contrast with a transitional disk consisting of a halo around the star, a dust gap and an hydrostatic outer disk, we successfully fit all the observations. 
\item Radiative transfer modelling constrains the inner radius of the outer disk at $20^{+3}_{-3}$ AU for HD~100453 and $92^{+31} _{-17}$ AU for HD~34282. This result provides further evidence that group Ib Herbig Ae/Be might have dust gap. Therefore, it supports the conclusion by \citet{2013Maaskant} that group Ib Herbig Ae/Be stars are transitional disks with large dust gaps. 
\item The upper limit for the mass of the dust in the gap is roughly estimated to be 10$^{-7}$M$_{\odot}$ for both HD~34282 and HD~100453.
\item The outer disk mass, surface density power law, and  the opacity profile do not affect the Q-band size. 
\item We find no correlation between the halo masses,  the disk masses and the sizes of the dust gaps in studied Herbig Ae/Be stars.
\item The absence of the amorphous silicate emission in the spectra of these disks is consistent with the conclusions of \citet{2013Maaskant} that dust gaps in a critical temperature regime between $\sim200-500$ K cause the silicate feature to disappear. In addition, grains are typically large in the outer disk.
\item The temperature transition from $\sim$160 K to $\sim$200 K in the inner edge of the outer disk is critical for the strong enhancement of the amorphous silicate features originating in the wall
\end{itemize}

\begin{acknowledgements}

The authors thank Rens Waters for inspiring discussions which helped to improve the analysis in this paper. The authors thank Michiel Min for providing his radiative transfer code MCMax. K.M. is supported by a grant from the Netherlands Research School for Astronomy (NOVA).  
Studies of interstellar chemistry at Leiden Observatory are supported through advanced-ERC grant 246976 from the European Research Council, through a grant by the Dutch Science Agency, NWO, as part of the Dutch Astrochemistry Network, and through the Spinoza premie from the Dutch Science Agency, NWO. 

 \end{acknowledgements}
 
\bibliographystyle{aa}
\bibliography{bib.bib} 

\begin{appendix}
\section{Inner disk parametrization}
\label{sec:appendix_inner}
Variations around the vertical scale-height $sh\propto r^{shpow}$ where $sh$ is the scaleheight at 0.1 AU and $shpow$ the power-law index. Models are performed on HD~100453. Our best efforts to fit the inner disk results in a poor fit to the outer disk. Thus, in our final model, we included a optically thin halo to solve for this problem. It should be mentioned that compared to the previous values shown in table 4, mass of halo is changed to $0.7 \times 10^{-1}$ M$_{\odot}$ and the extension of the inner disk is changed to 0.25$-$1.7 AU.

\begin{figure*}[ht]
	\centering
\includegraphics[width= 0.48\textwidth]{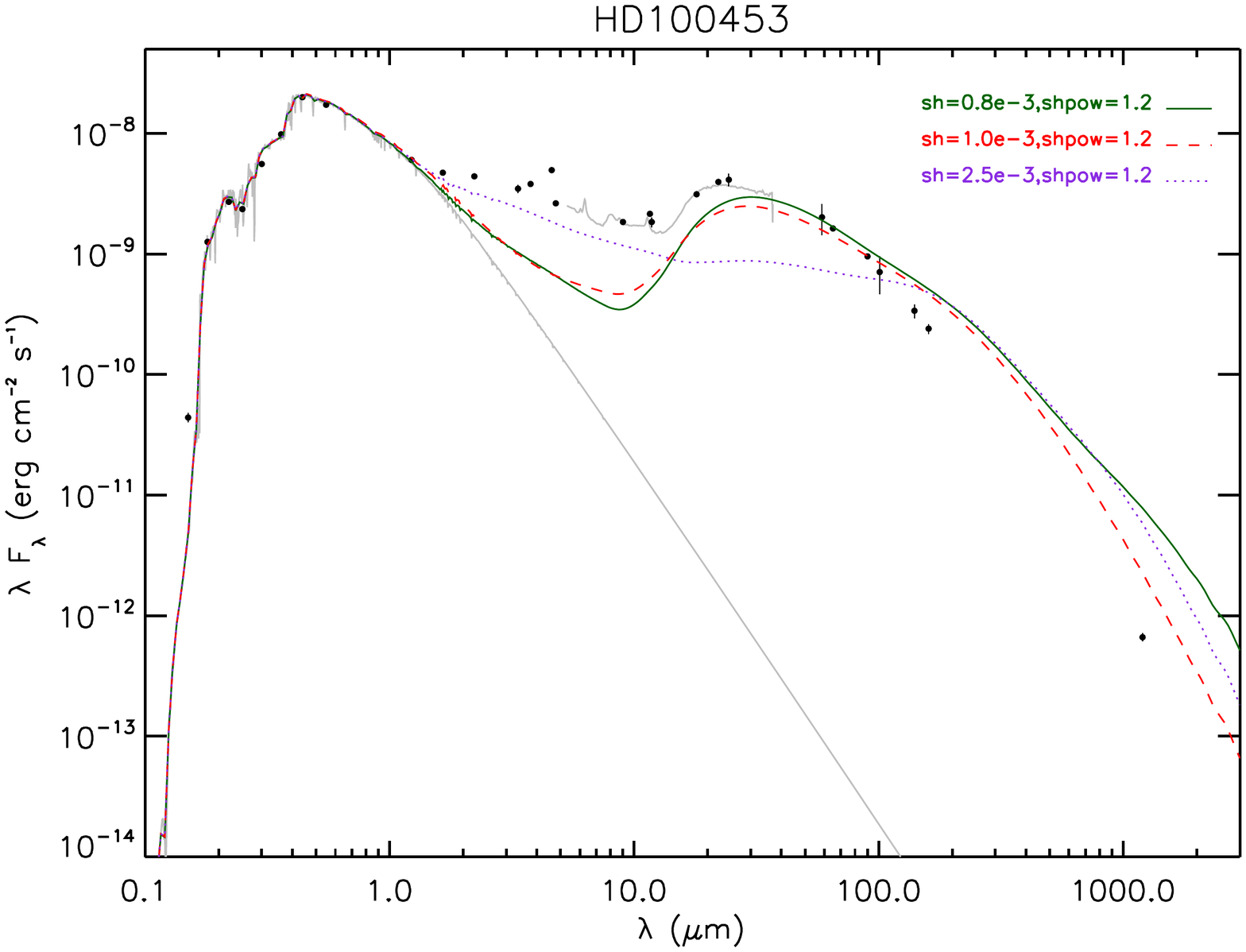}
\includegraphics[width= 0.49\textwidth]{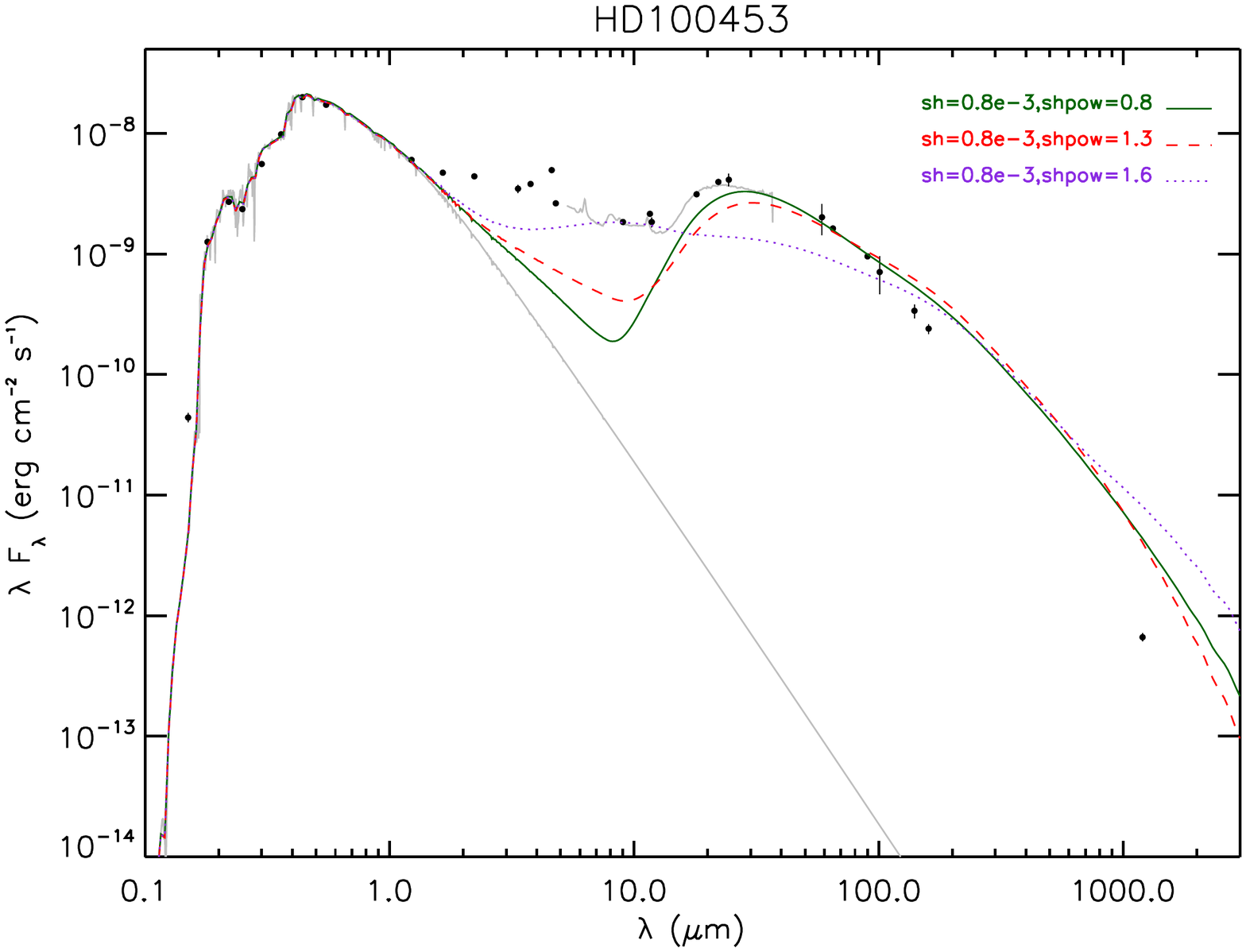}

\caption{\label{innerdisk_hd100453} Variations in the inner disk scale-height properties. Left: different scale-heights. Right: different power-law indices. No satisfactory solution using an optically thick inner disk could be found.}
\end{figure*}

\section{parameter study}
\label{sec:appendix}
\begin{figure*}[ht]
	\centering
\includegraphics[width= 0.48\textwidth]{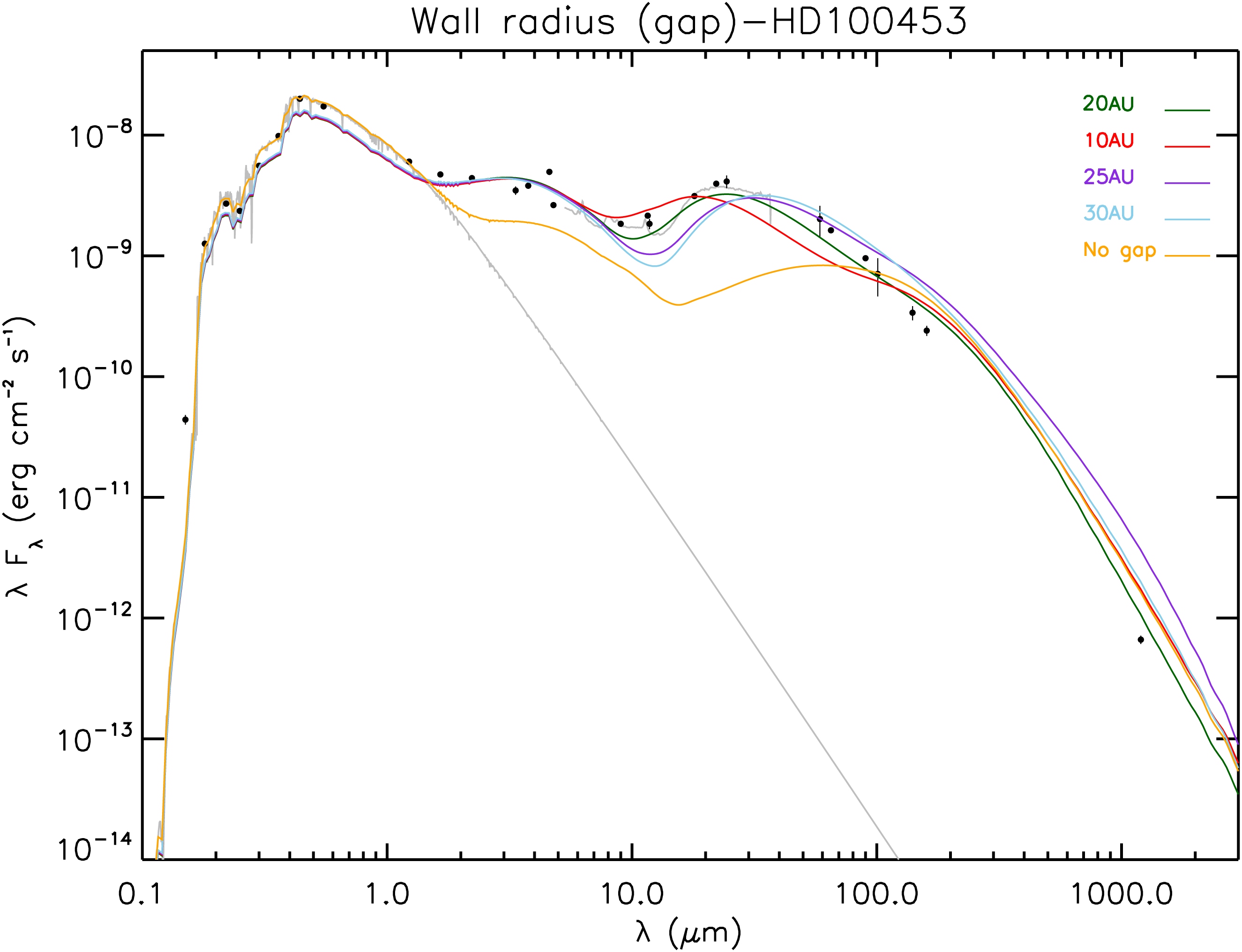}
\includegraphics[width= 0.49\textwidth]{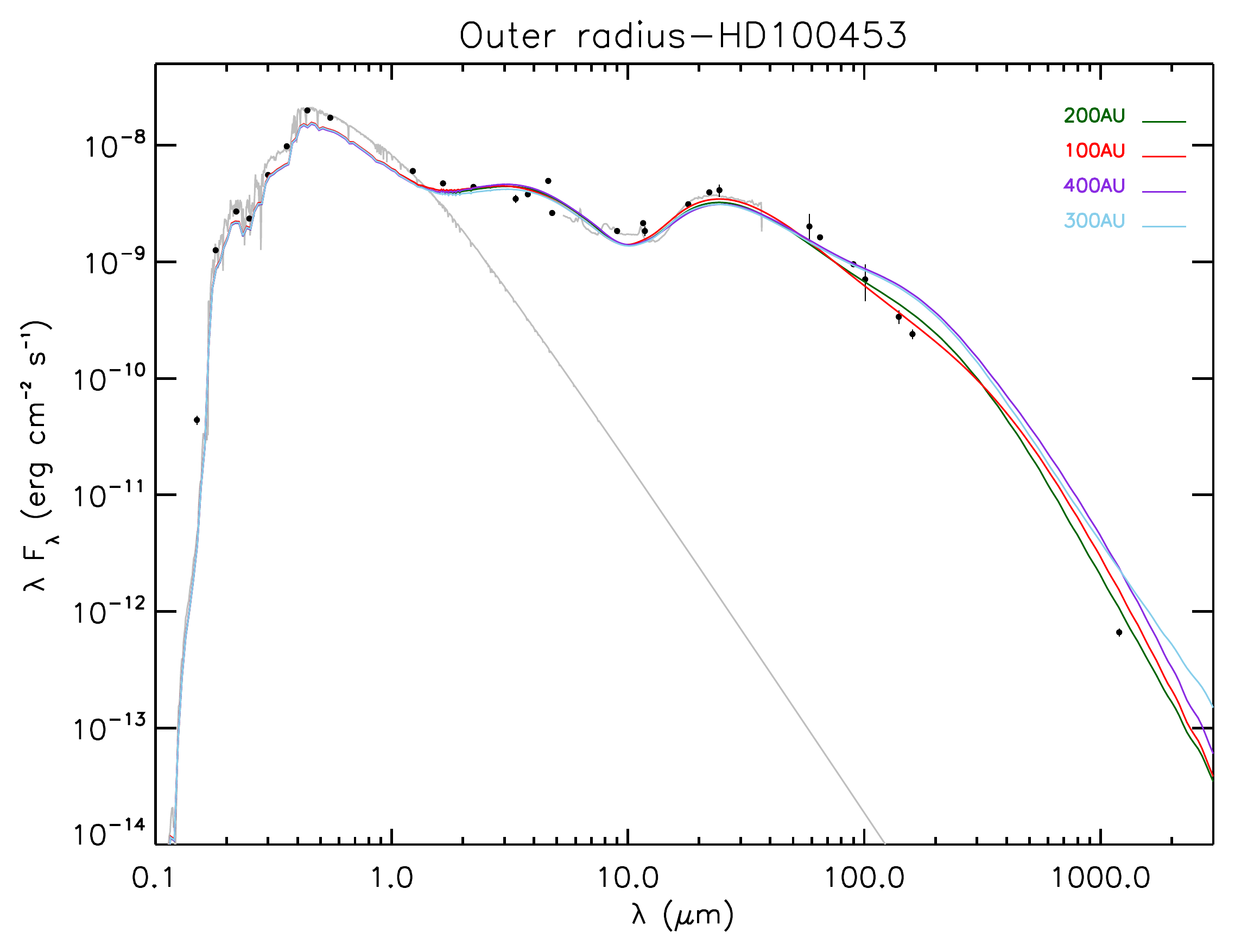}
\includegraphics[width=0.49\textwidth]{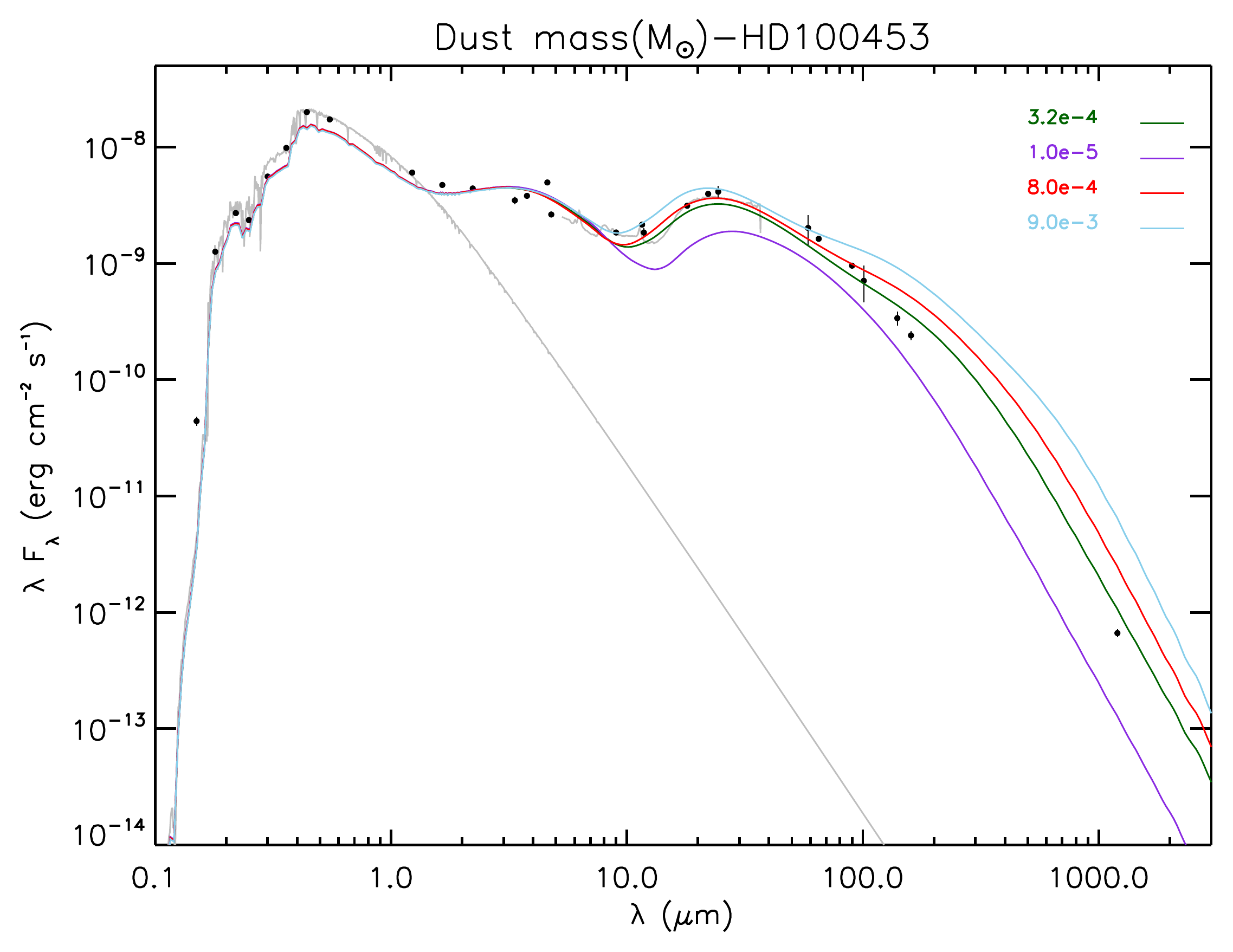}
\includegraphics[width= 0.49\textwidth]{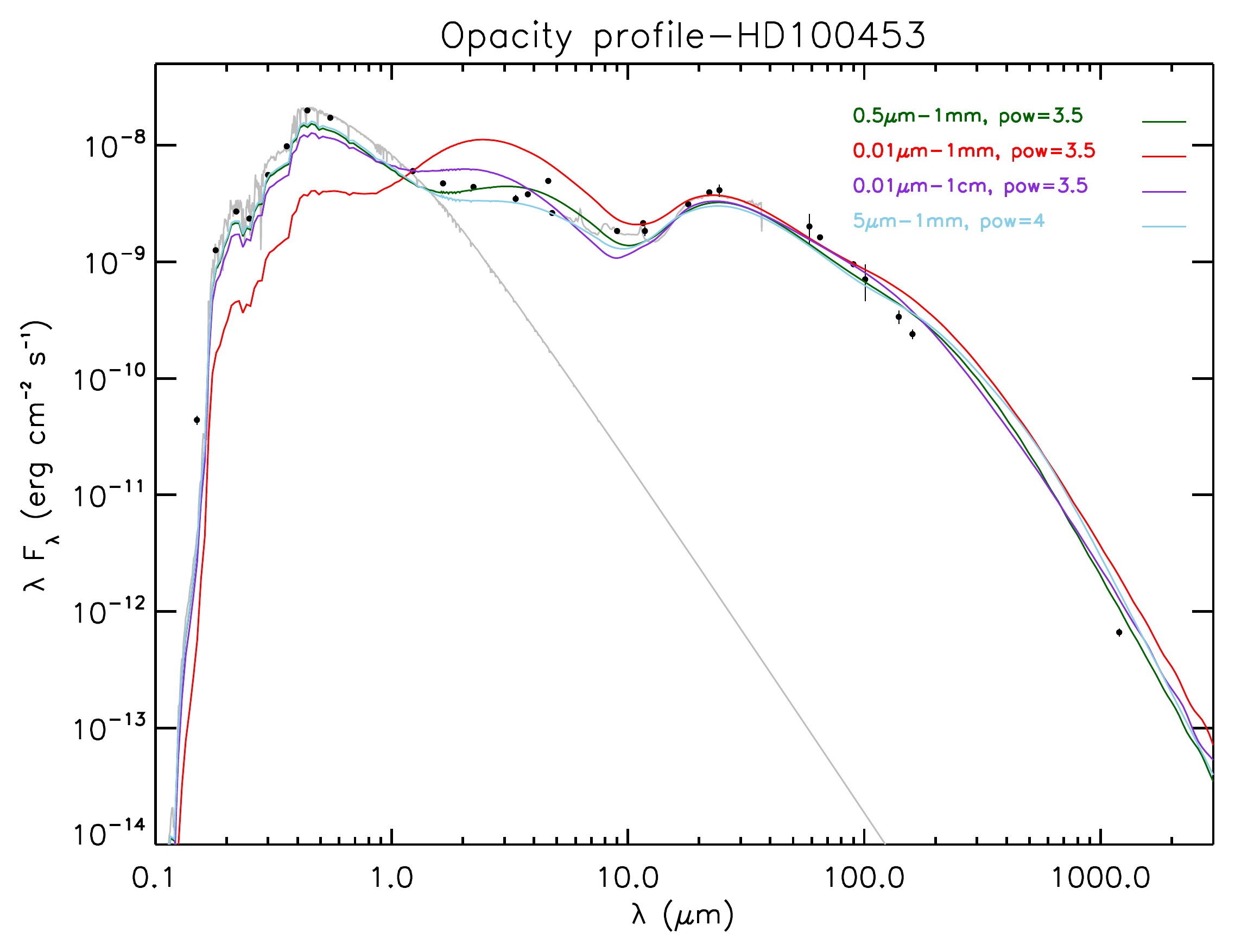}

\caption{\label{SEDhd100453_1} Variations around the best-fit values for HD~100453. Top left: different dust gap sizes. Top right: different radii of outer disk. Bottom left: different total disk masses. Bottom right: different minimum and maximum grain sizes and power-laws of the grain size distribution.}
\end{figure*}

Figure \ref{SEDhd100453_1} shows the results of a small parameter study to illustrate the degeneracy of SED modeling. HD~100453 is shown on the left and HD~34282 is shown on the right.

The top left plot in Figure \ref{SEDhd100453_1} show the result of modelling different values for the inner radius of the outer disk (the wall radius). This value affects the MIR of the SED where the emission comes from the wall and the temperature is about $\sim$100-160 K. The dust gap size with the value of 20 AU fits the observed SEDs very well. This indicates a presence of a dust gap in between the halo and the outer disk. The exact size of this dust gap cannot be understood from modeling the SED alone, but must include fitting the size of the Qband image. The SED is degenerate because the structure of the inner halo (i.e. the optical depth, and scaleheigth), as well as the grain composition in the outer disk  are not well known. As we will see, these parameters have a great influence on the SED as well. 

The top right plot in Figure \ref{SEDhd100453_1}, shows a comparison of different disk radii. The outer radius is unknown and we try some values in the range of typical disk sizes and we find that a disk size with radius of 200 AU fits better the SED better. However, the changes of this value do not have a significant effect on the outcome of the modelling and therefore, it is not easy to estimate the size of the outer radius of the disk.

The bottom left plot in Figure \ref{SEDhd100453_1} show variations in the dust mass. Higher mass means more material in the disk and therefore this would increase the thermal emission from the disk in the far-IR regions. Dust mass mainly affects the mm part of the SED. When the dust mass is taken to be lower than the best-fit value, the mm flux tends to shifts below the observation and when the mass is higher than the best fit, the mm flux shifts above the observation.

The bottom right plot in Figure \ref{SEDhd100453_1} shows the result for different opacity profiles (i.e. different grain sizes and power law indices). The best-fit consists of grain size distribution of 0.5 $\mu$m to 1 mm with the power law index of -3.5. The change in this parameter affects the temperature and luminosity of the halo and the location of the wall of the outer disk. This is especially noticeable in the NIR part of the SEDs. The red lines show the models with the highest abundance of small grains. Smaller grains have higher opacities, this means that more stellar radiation is absorbed the close to the star. Therefore the emission from the halo increases, but the stellar radiation decreases due to the higher optical depth. The other grain size populations show similar behaviours.

\begin{figure*}
	\centering
\includegraphics[width= 0.43\textwidth]{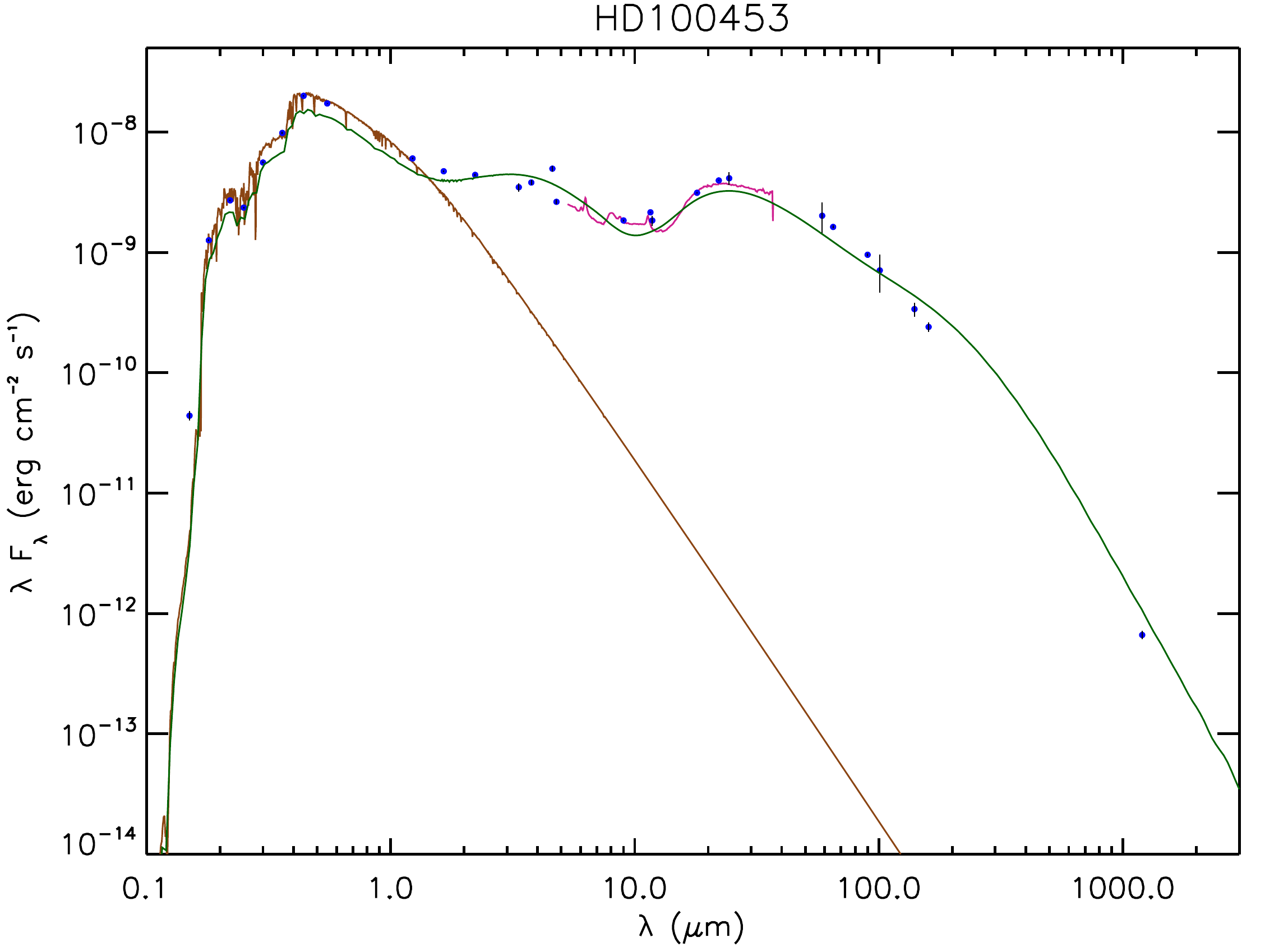}
\includegraphics[width= 0.44\textwidth]{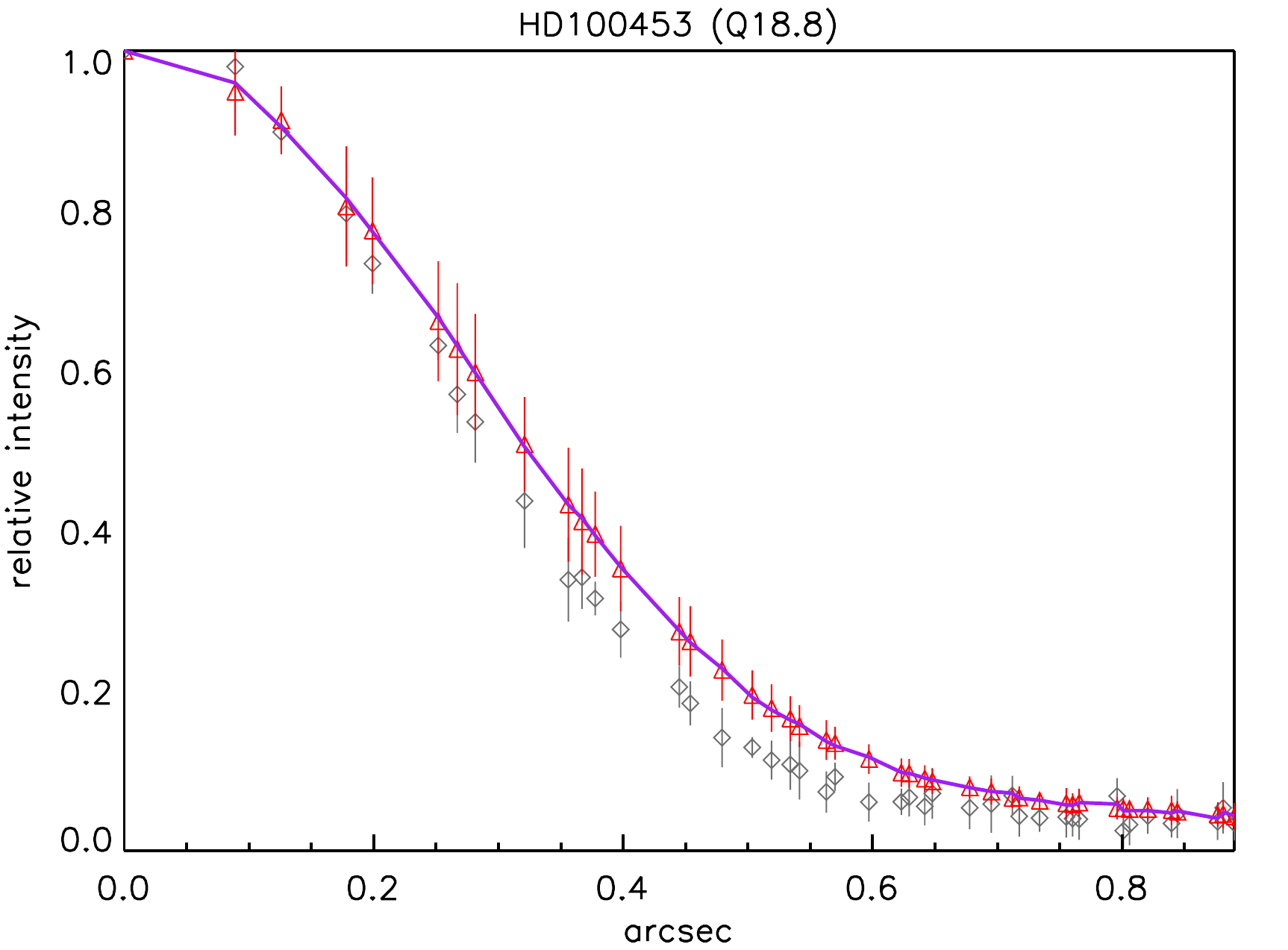}
\includegraphics[width= 0.43\textwidth]{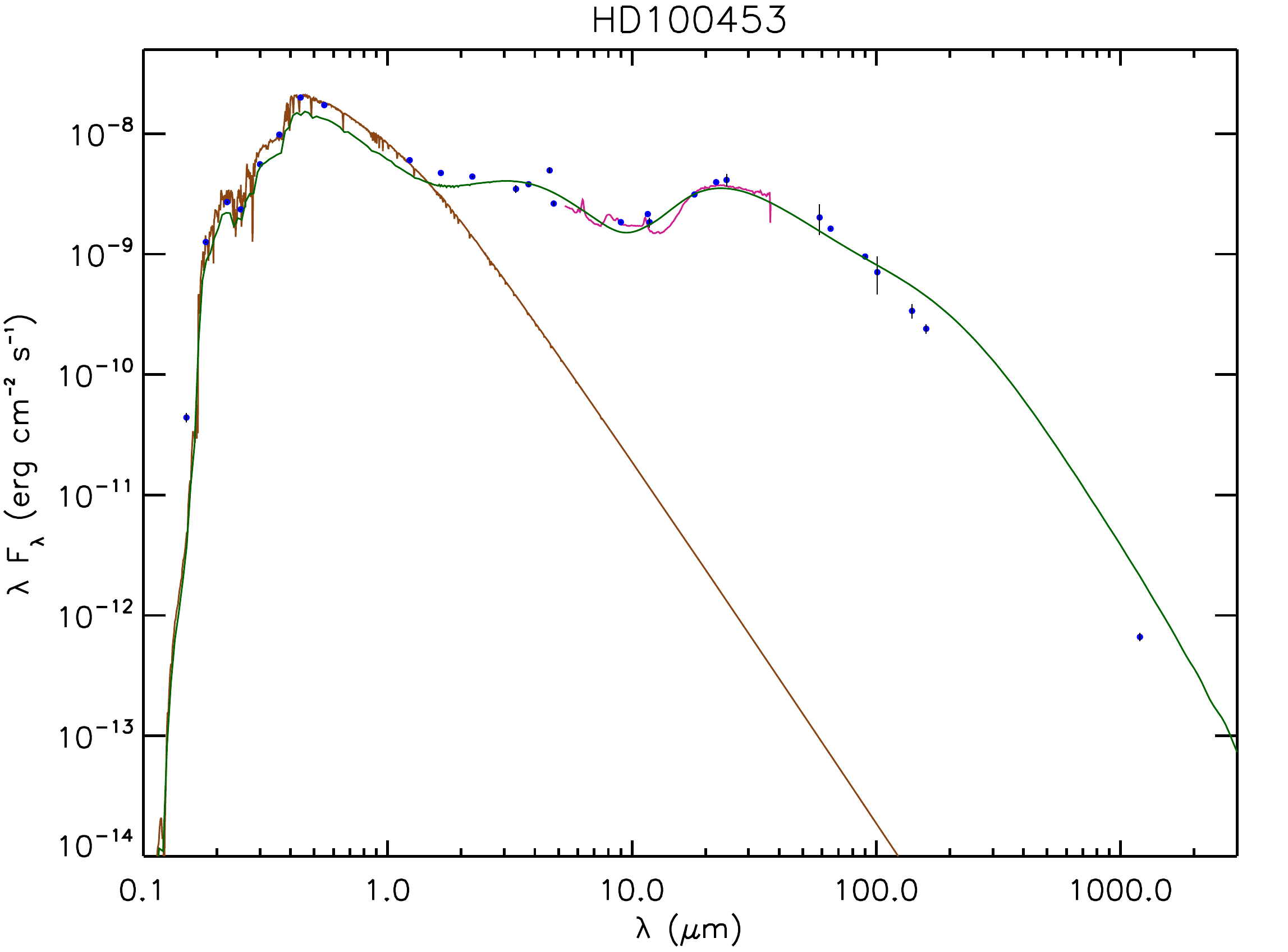}
\includegraphics[width= 0.44\textwidth]{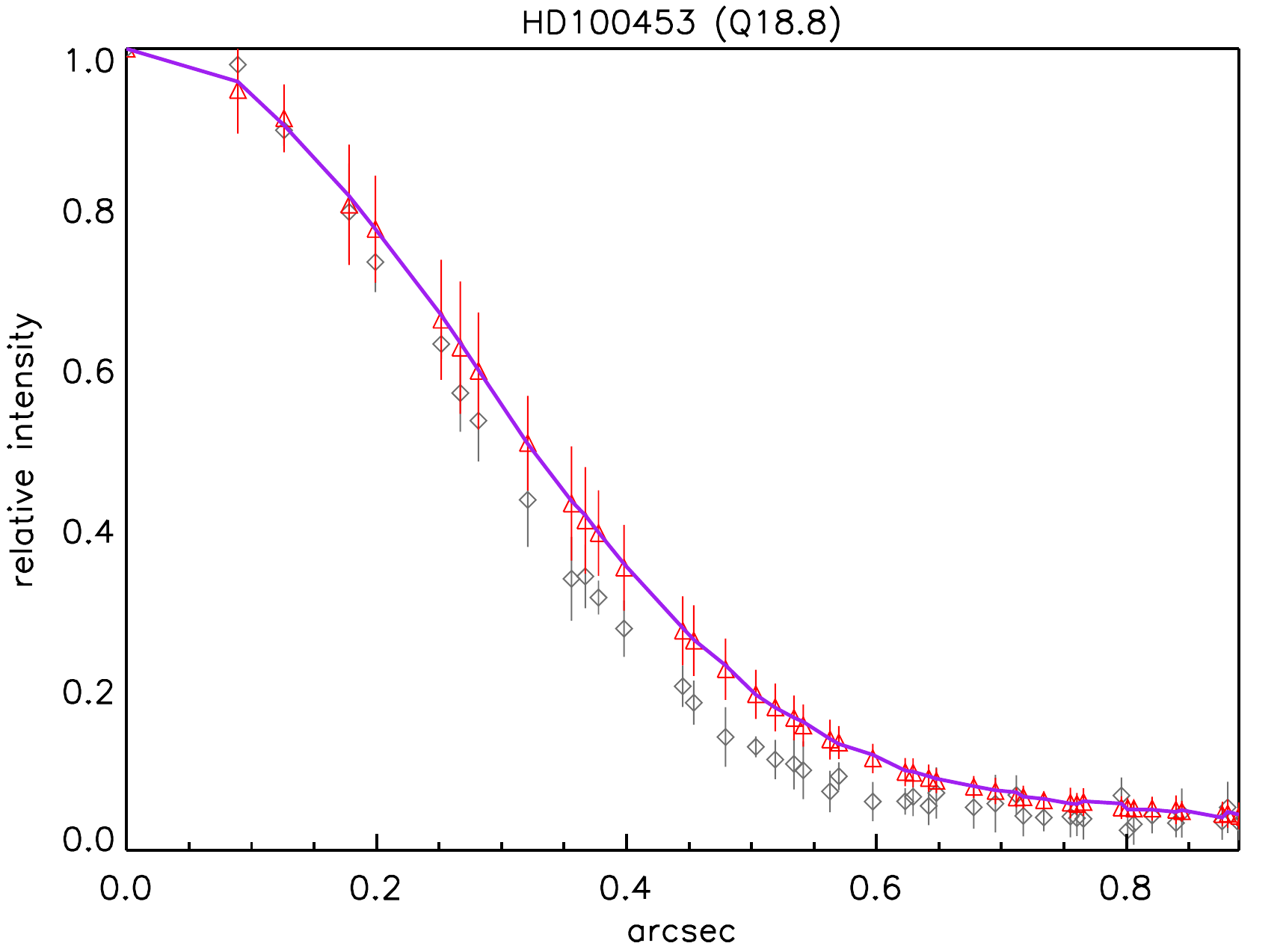}
\includegraphics[width= 0.43\textwidth]{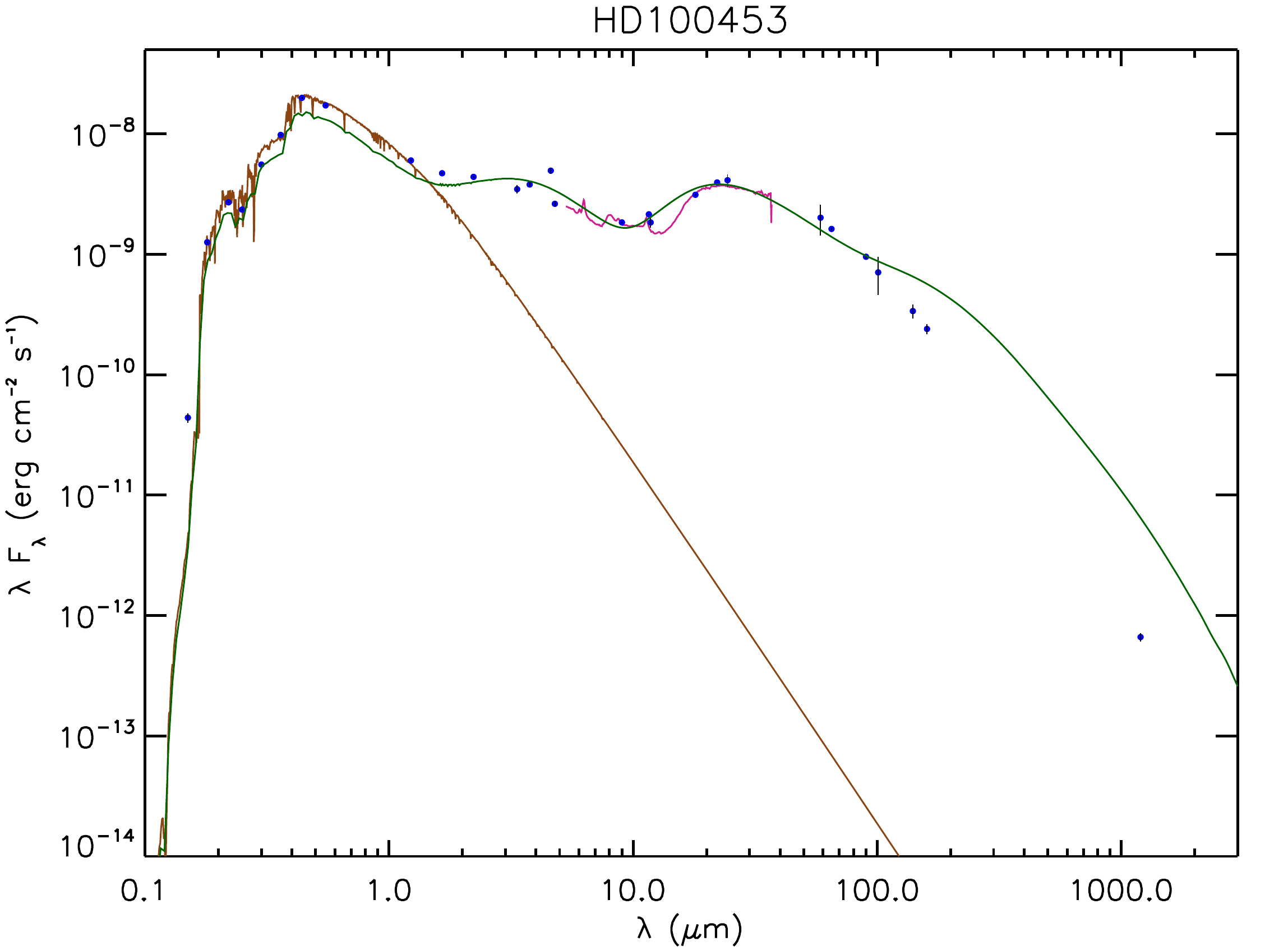}
\includegraphics[width= 0.44\textwidth]{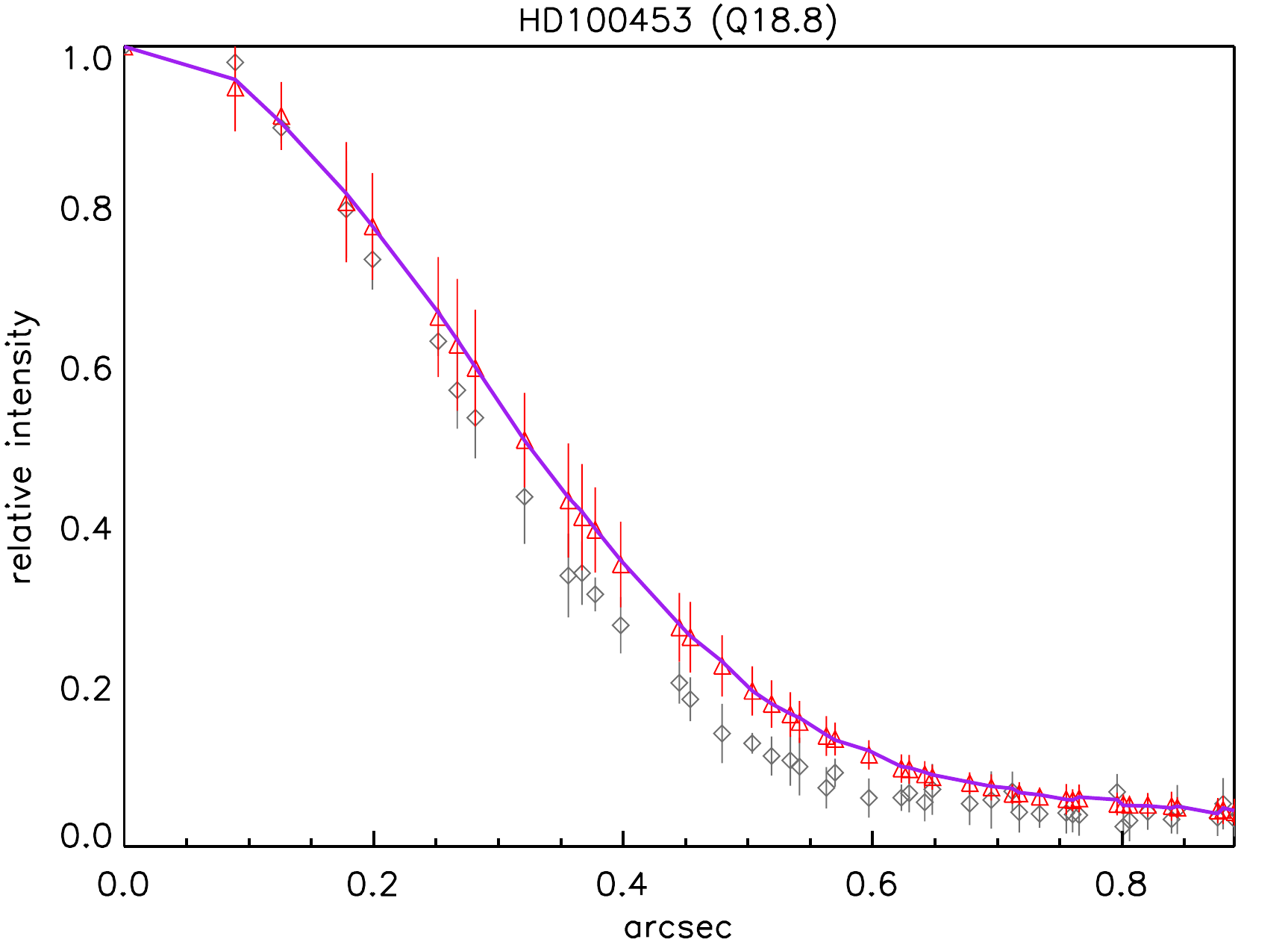}
\includegraphics[width= 0.43\textwidth]{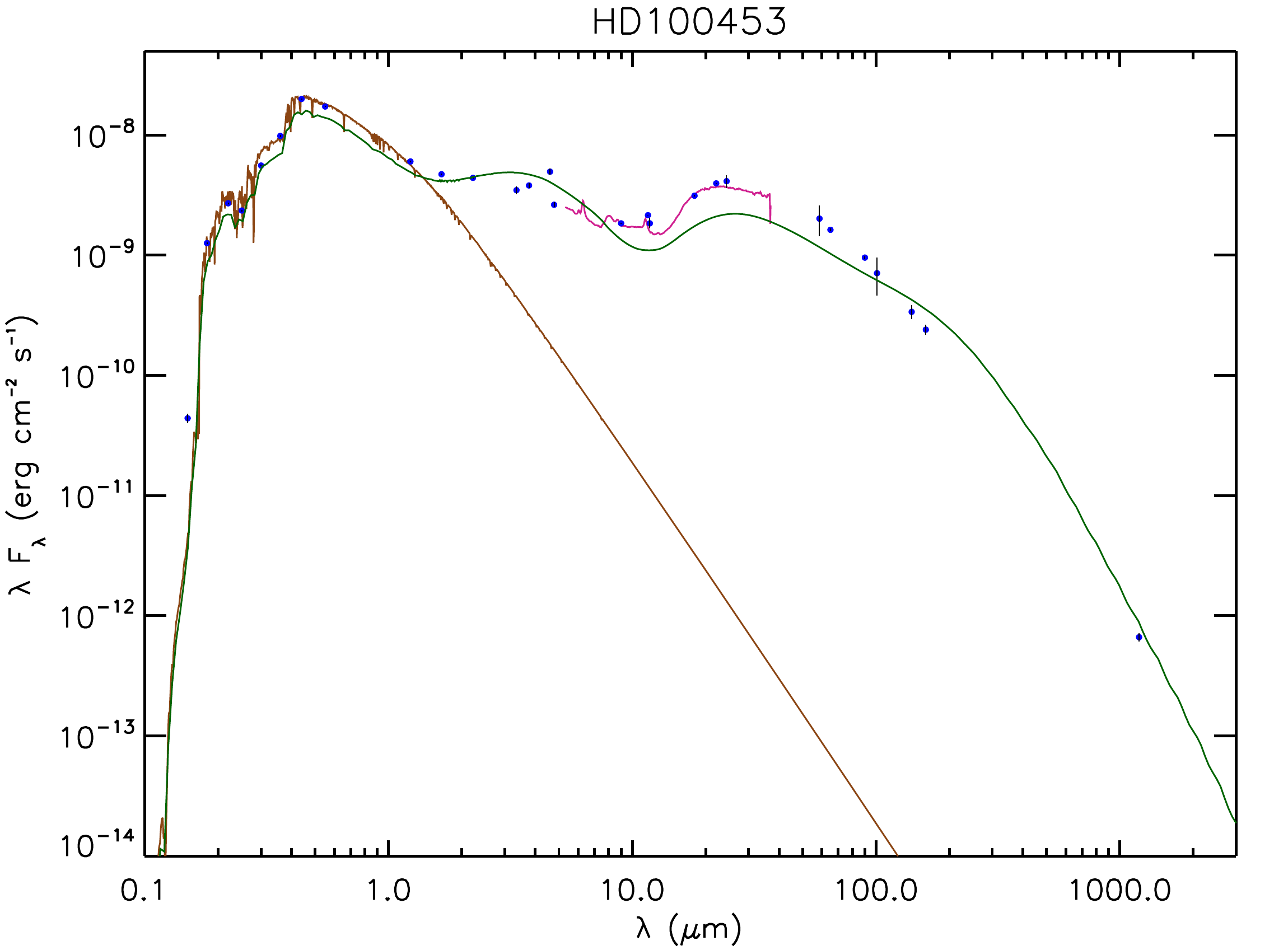}
\includegraphics[width= 0.44\textwidth]{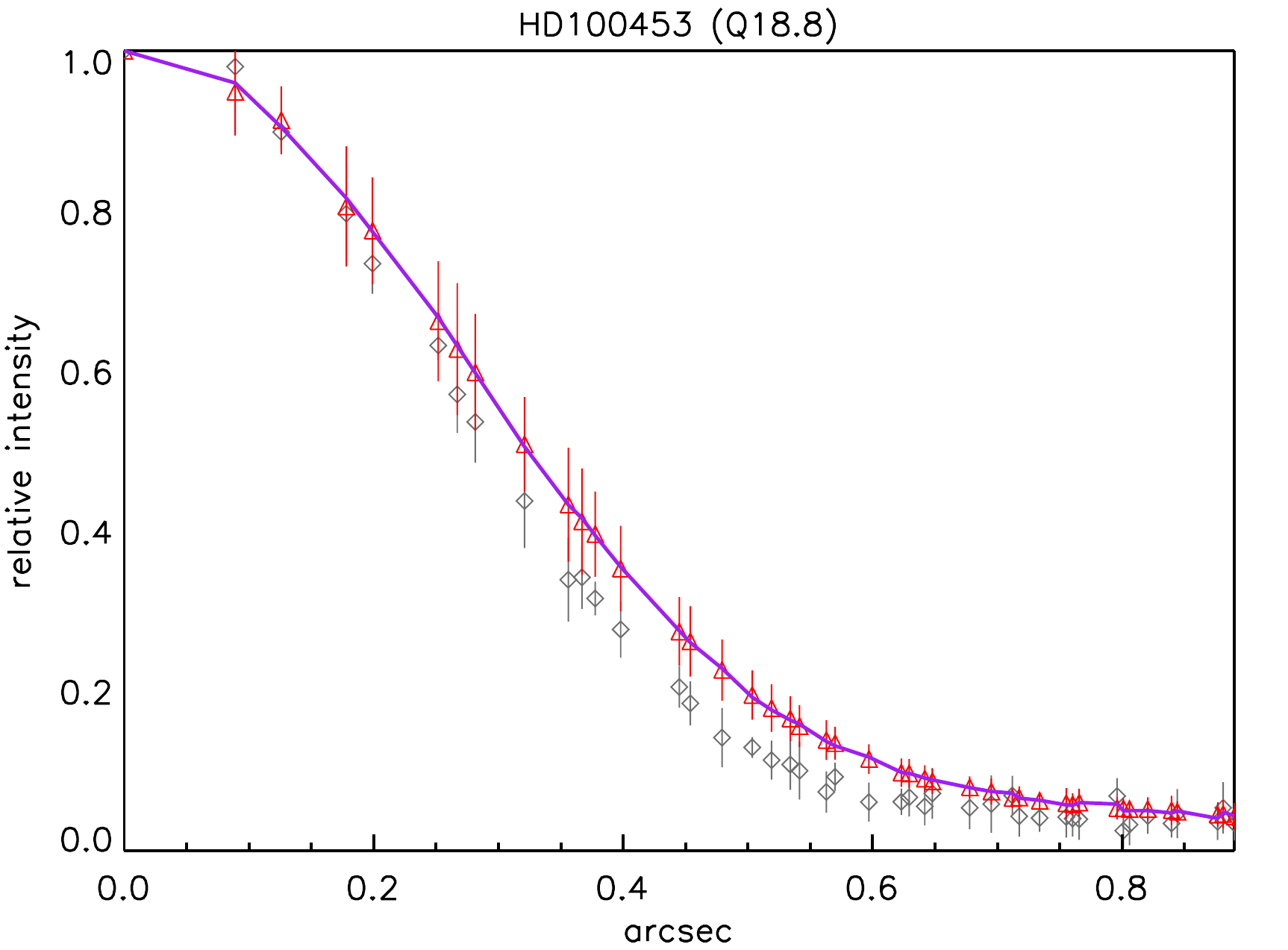}

\caption{\label{compvssize} The SED and RBP for different compositions. First row shows SED and RBP for a model with 80\% carbon and 20\% silicates. Second row shows a model with 50\% carbon and 50\% silicates, third row shows a model with 20\% carbon and 80\% silicates and the fourth row shows a model with 100 \% carbon and no silicates.}
\end{figure*}

The plots on Figure \ref{compvssize} show the SED and Radial brightness profiles for several compositional ratios of carbons to silicates. While there is some change in the shape of the SED, little difference can be observed in the size of the Q-band image.

\section{Supplementary to the upper limit on the dust mass in the gap}
\label{upper appendix}

\begin{figure*}
	\centering

\includegraphics[width=0.4\textwidth]{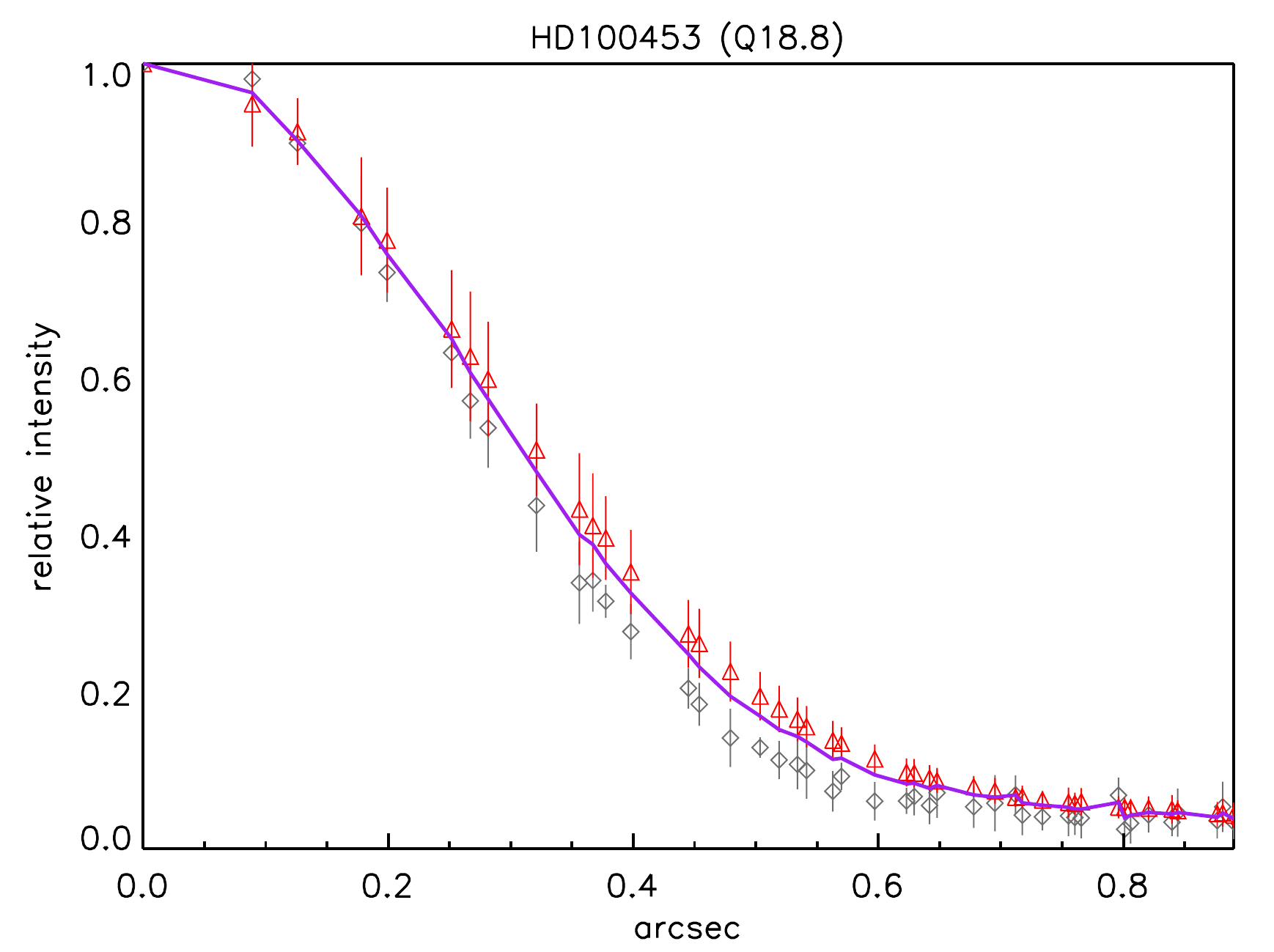}
\includegraphics[width=0.4\textwidth]{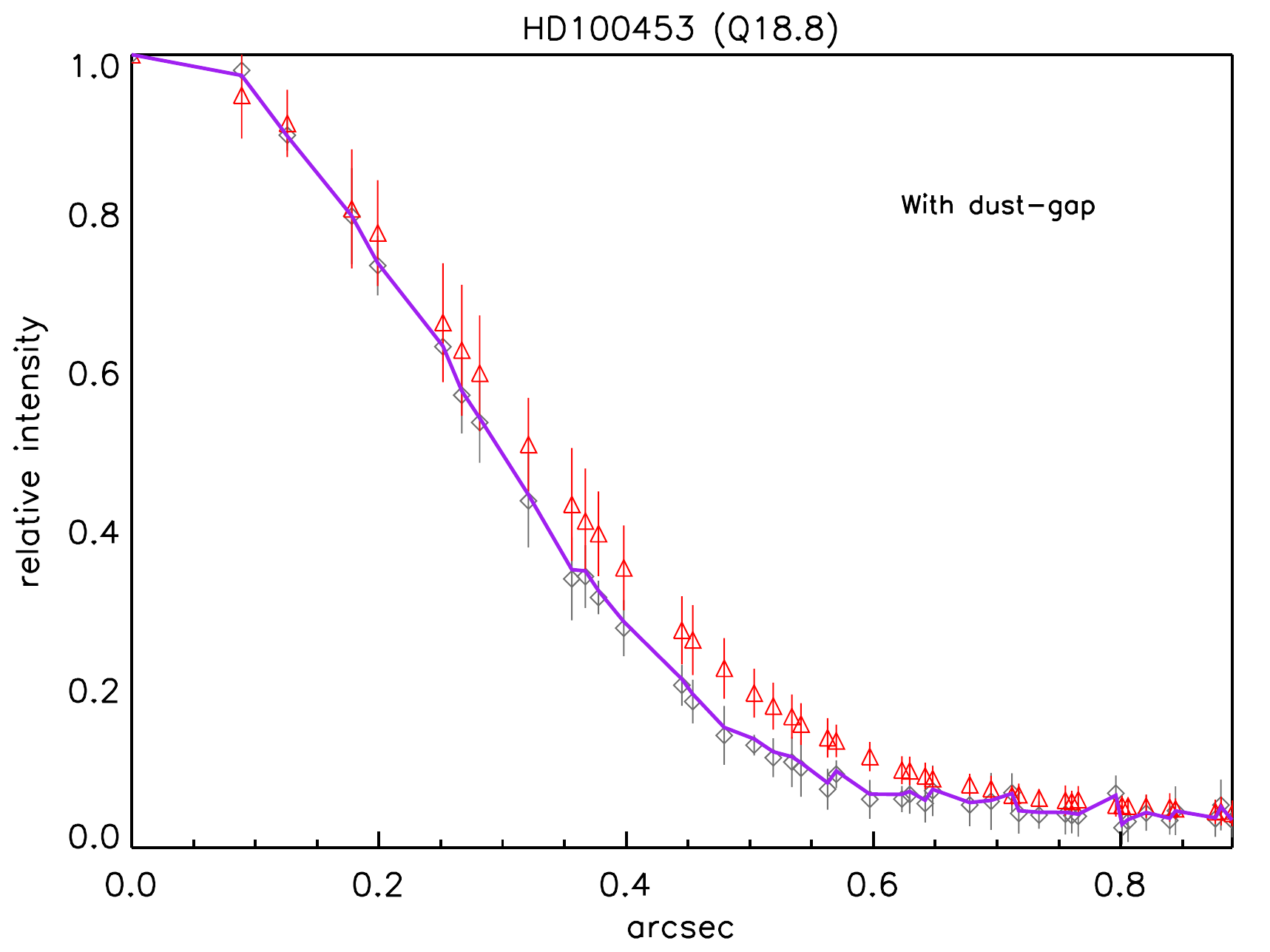}
\includegraphics[width=0.4\textwidth]{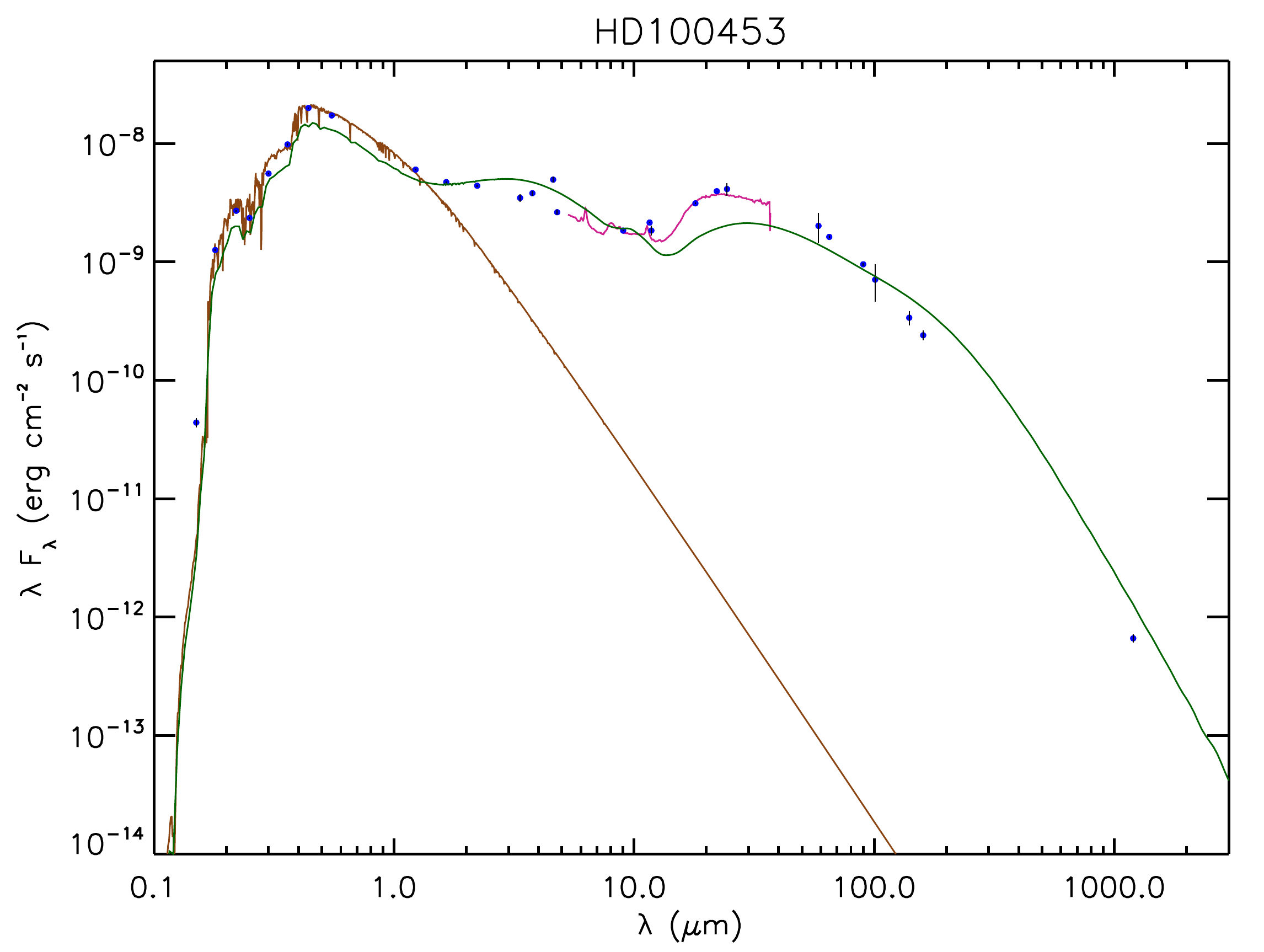}
\includegraphics[width=0.4\textwidth]{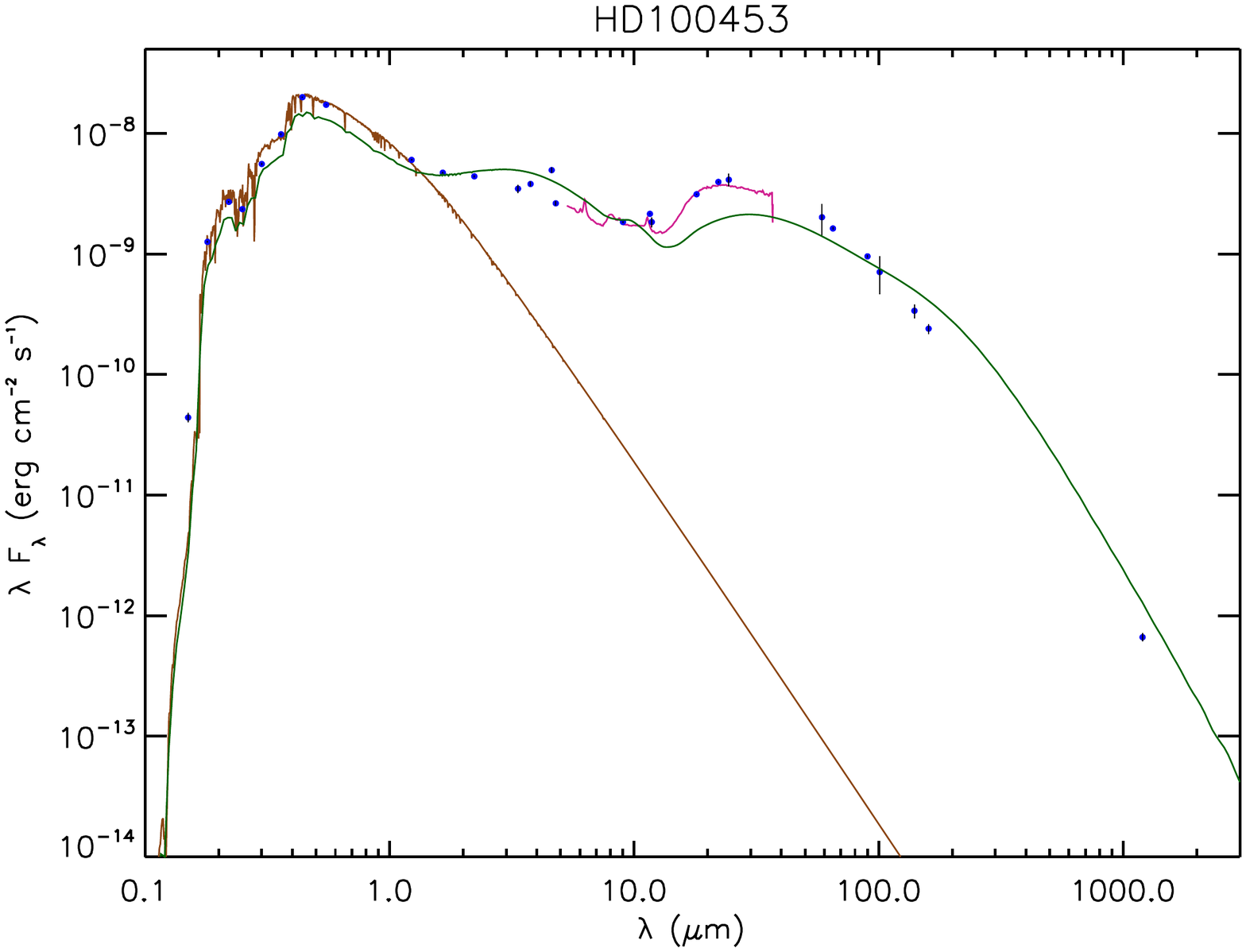}

\caption{\label{upper limit_HD100 appendix} Top panels show the radial brightness profiles of HD~100453 for when the mass of the dust in the gap is 10$^{-7}$M$_{\odot}$ and 10$^{-5}$M$_{\odot}$ from left to right respectively. Bottom panels show the corresponding SEDs.}
\end{figure*}

\begin{figure*}
	\centering

\includegraphics[width=0.4\textwidth]{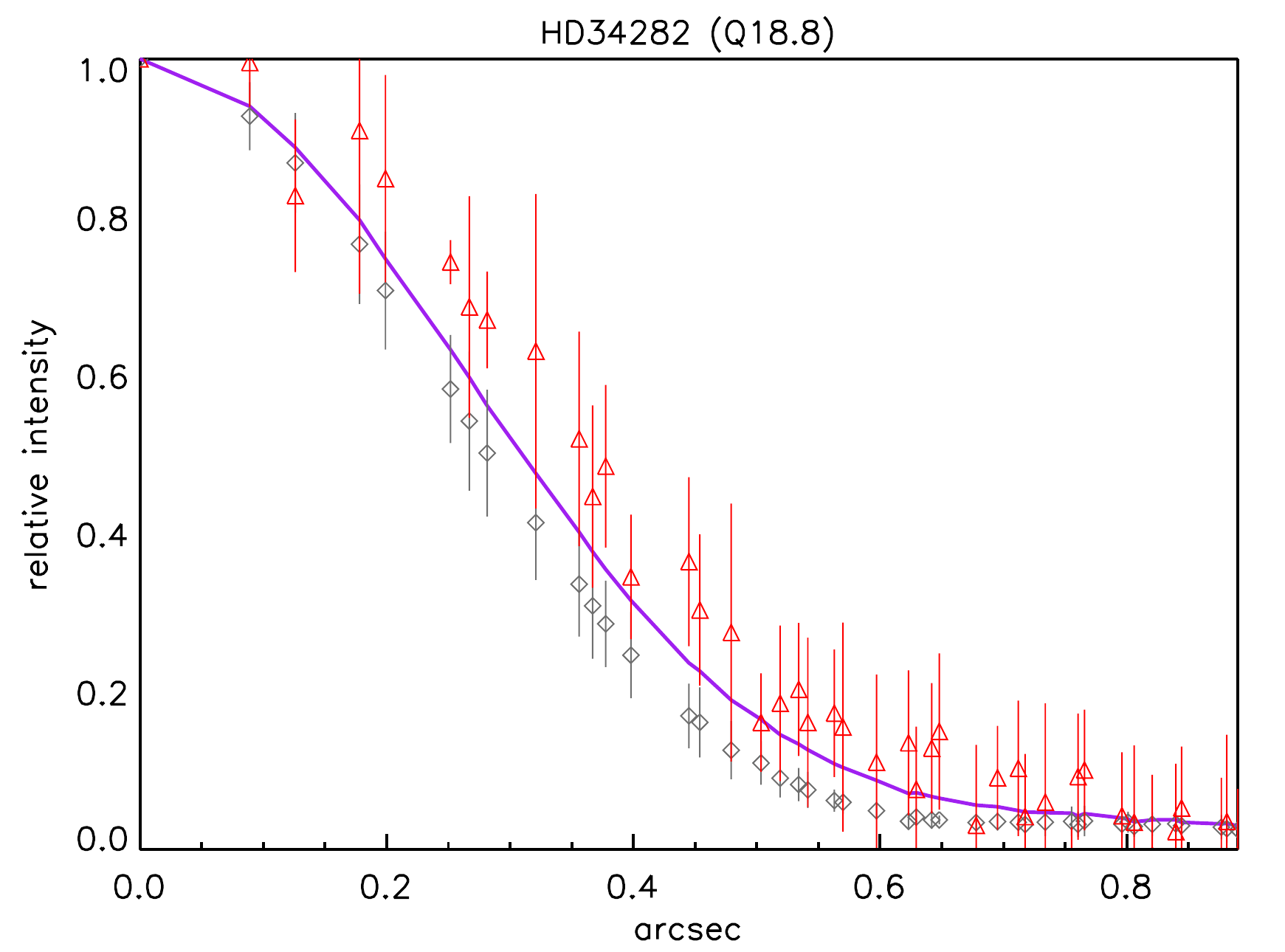}
\includegraphics[width=0.4\textwidth]{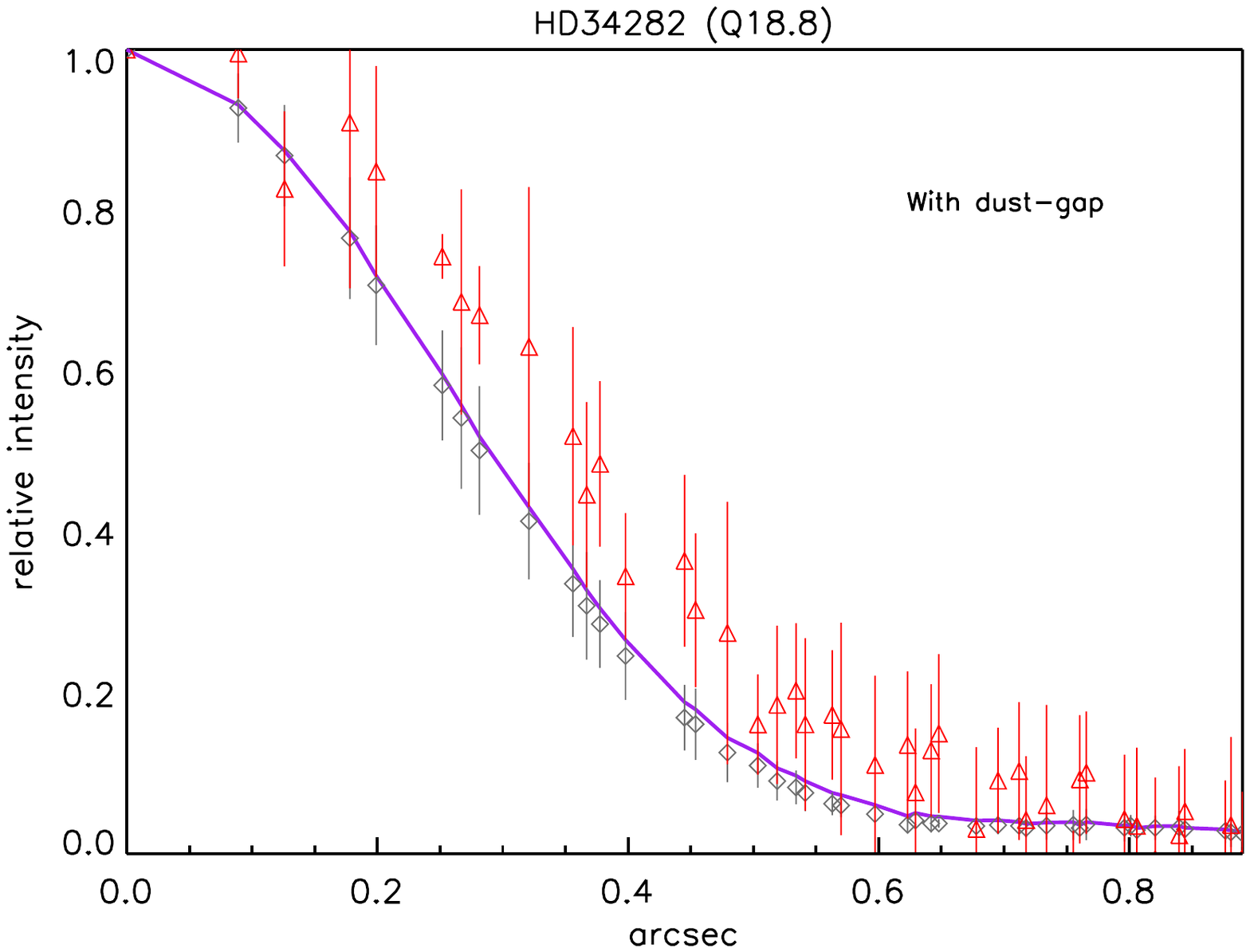}
\includegraphics[width=0.4\textwidth]{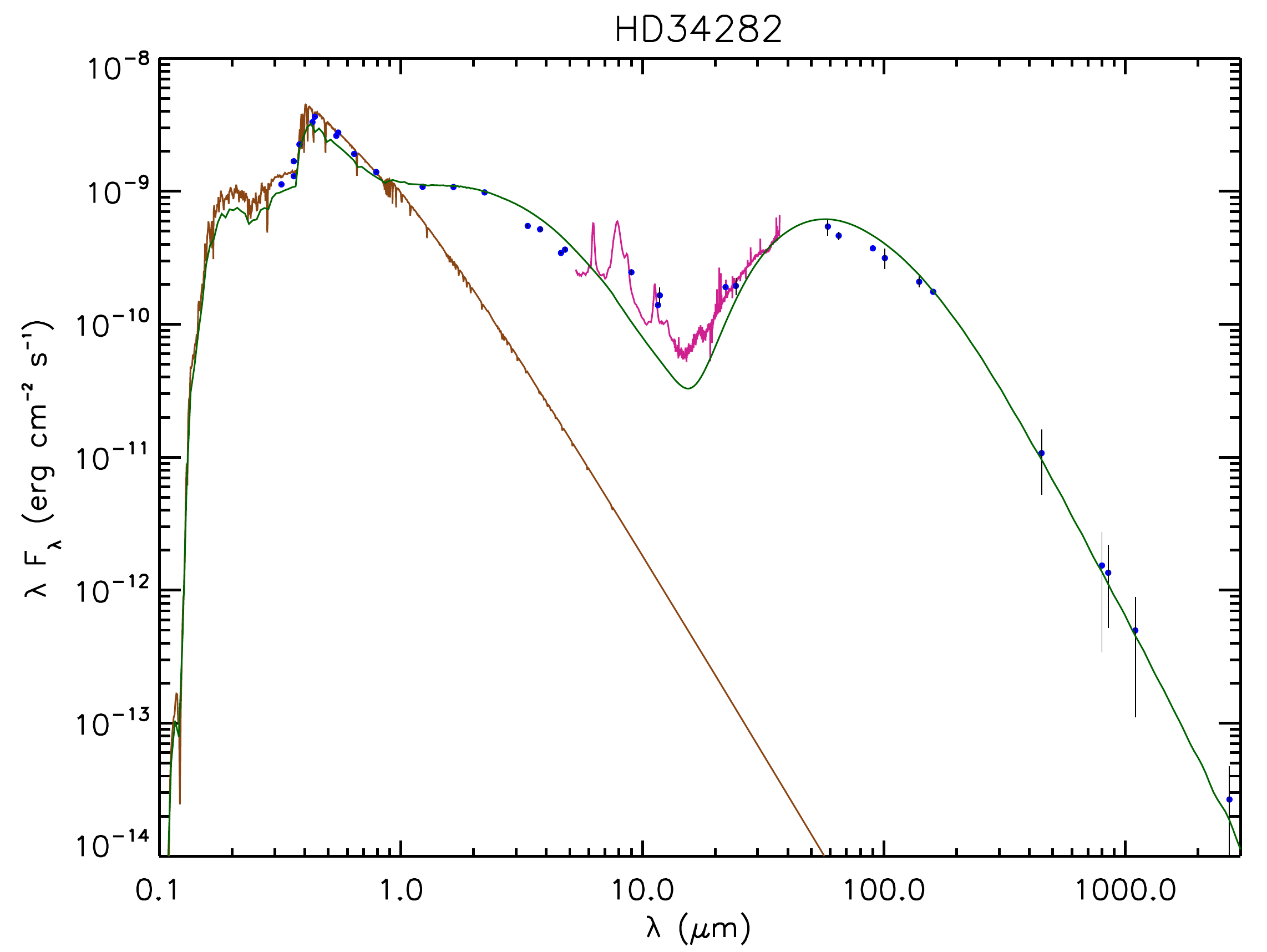}
\includegraphics[width=0.4\textwidth]{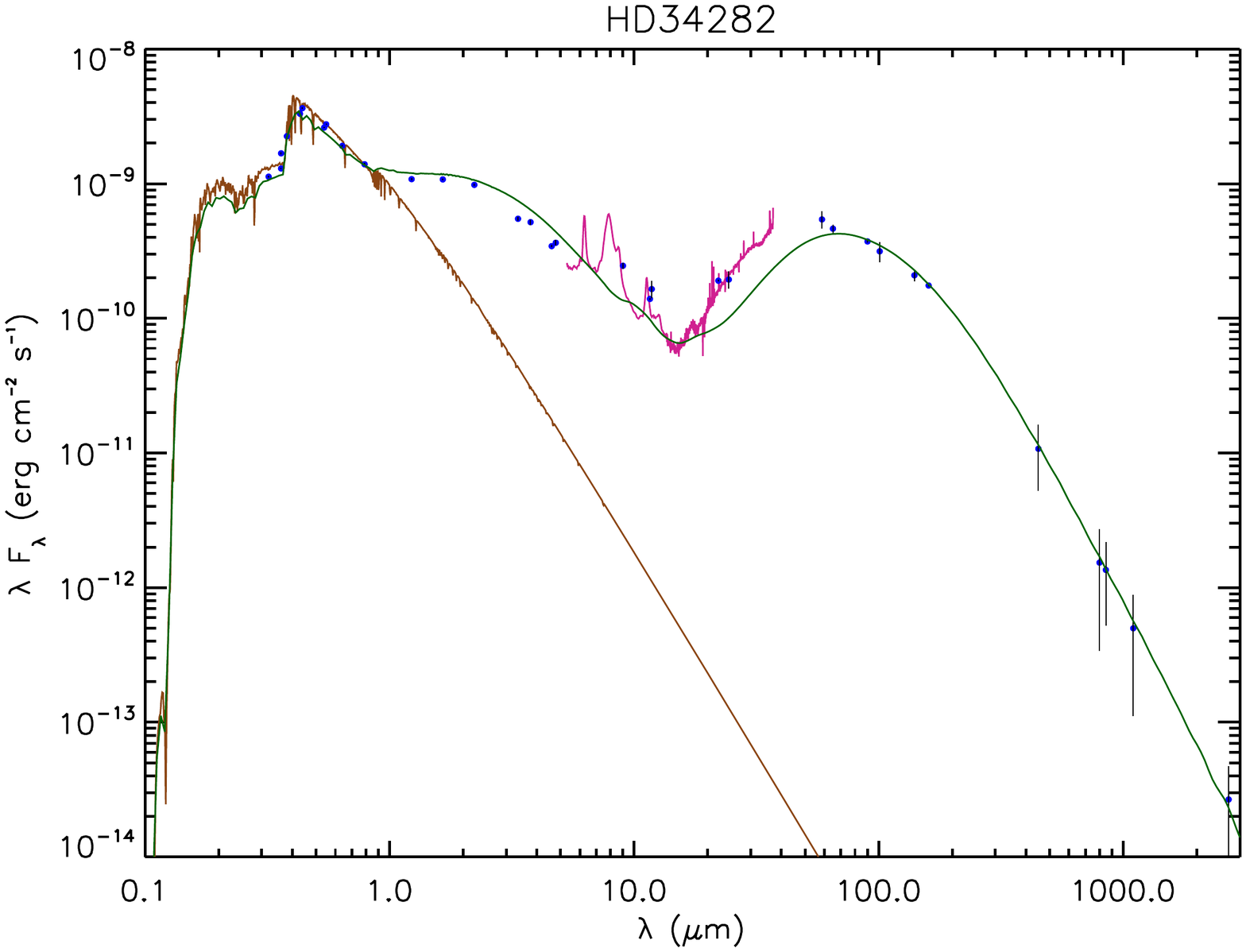}

\caption{\label{upper limit_HD342 appendix} Top panels show the radial brightness profiles of HD~34282 for when the mass of the dust in the gap is 10$^{-7}$M$_{\odot}$ and 10$^{-6}$M$_{\odot}$ from left to right respectively. Bottom panels show the corresponding SEDs.}
\end{figure*}

As mentioned in section \ref{sec:upper limit}, we can estimate the upper limit on the dust mass inside the gap by increasing the dust mass in the gap. In this region for masses above $\sim10^{-7}$M$_{\odot}$ for both sources, the observed size of the image in the Q-band cannot be reproduced anymore (shown in the top left panels in Figures \ref{upper limit_HD100 appendix} and \ref{upper limit_HD342 appendix}). In addition, for comparison, we have demonstrated the SED and RBP of both sources for even higher masses in the right panels of Figures \ref{upper limit_HD100 appendix} and \ref{upper limit_HD342 appendix}.
The main effect of an increased dust mass in the gap is that the inner region becomes optically thick. As a
result the wall (inner radius of the outer disk) becomes less pronounced and therefore less important for the SED and
for the Q-band brightness distribution. This effect is readily apparent in
the models shown in Figures \ref{upper limit_HD100 appendix} and \ref{upper limit_HD342 appendix}.

\end{appendix}

\end{document}